%% file: main.tex
\documentclass[%
reprint,
superscriptaddress,
 amsmath,amssymb,
aps,
prb,
]{revtex4-2}
\usepackage{soul}
\usepackage{pifont}
\usepackage{verbatim}
\usepackage{subcaption}
\usepackage{graphicx}%
\usepackage{dcolumn}%
\usepackage{bm}%
\usepackage{xcolor}
\usepackage[hidelinks]{hyperref}%
\usepackage{url}
\usepackage[normalem]{ulem} %

\usepackage{physics}
\usepackage{amsthm,amsbsy,amsfonts,gensymb}
\usepackage{wrapfig}

\newcommand{\Z}{\mathbb{Z}}

\newcommand{\R}{\mathbb{R}}

\newcommand{\I}{\mathbb{I}}

\usepackage[utf8]{inputenc}
\usepackage{pgfplots}
\DeclareUnicodeCharacter{2212}{−}
\usepgfplotslibrary{groupplots,dateplot}
\usetikzlibrary{patterns,shapes.arrows,math}
\pgfplotsset{compat=newest}
\usepackage{tikz-3dplot}

\usepackage{xr,cleveref}
\makeatletter
\newcommand*{\addFileDependency}[1]{
    \typeout{(#1)}
    \@addtofilelist{#1}
    \IfFileExists{#1}{}{\typeout{No file #1.}}
}\makeatother

\newcolumntype{P}[1]{>{\centering\arraybackslash}p{#1}}
\newcolumntype{M}[1]{>{\centering\arraybackslash}m{#1}}

\iftrue
\usepackage{pdfpages}
\usepackage{pgffor}
\makeatletter
\AtBeginDocument{\let\LS@rot\@undefined}
\makeatother
\fi

\begin{document}

\title{Intrinsically-multilayer moir\'e heterostructures}

\author{Aaron Dunbrack}
\affiliation{Department of Physics and Astronomy, Stony Brook University, Stony Brook, New York 11974, USA}

\author{Jennifer Cano}
\affiliation{Department of Physics and Astronomy, Stony Brook University, Stony Brook, New York 11974, USA}
\affiliation{Center for Computational Quantum Physics, Flatiron Institute, New York, New York 10010, USA}

\date{\today}

\begin{abstract}
    We introduce trilayer and multilayer moir\'e heterostructures that cannot be viewed from the ``moir\'e-of-moir\'e" perspective of helically-twisted trilayer graphene. These ``intrinsically trilayer" moir\'e systems feature periodic modulation of a local quasicrystalline structure. They open the door to realizing moir\'e heterostructures with vastly more material constituents because they do not constrain the lattice constants of the layers. In this manuscript, we define intrinsically multilayer patterns, provide a recipe for their construction, derive their local configuration space, and connect the visual patterns to physical observables in material systems.
\end{abstract}

\maketitle

\section{Introduction\label{Sec:Intro}}

\begin{table}
    \centering
    \begin{tabular}{|P{0.2\linewidth}|P{0.35\linewidth}|P{0.35\linewidth}|}
        \hline
        Types of moir\'e patterns & Small twist & Large twist
        \\ \hline Two layers
            & \include{Images_TableImages_TBLGSchematic} Twisted bilayer graphene
            & \include{Images_TableImages_CommensTBLGSchematic} Near-commensurate TBLG
        \\ \hline Three or more layers
            & \include{Images_TableImages_TTLGSchematic} Twisted trilayer graphene
            & \include{Images_TableImages_IntrinsicallyTrilayerSchematic} Intrinsically trilayer moir\'e
        \\ \hline
    \end{tabular}
    \caption{Summary of moir\'e heterostructures: the ``intrinsically trilayer'' moir\'e patterns we introduce occur at large twist angle and with three or more layers.}
    \label{tbl:IntroSummary}
\end{table}

The observation of superconductivity and correlated insulators in twisted bilayer graphene
\cite{Cao_2018,cao2018correlated}
launched the study of ``moir\'e materials,'' where two-dimensional materials with the same \cite{Cao_2018,cao2018correlated,Bistritzer12233,yankowitz2019tuning,lu2019superconductors,serlin2020intrinsic,nuckolls2020strongly,chen2020tunable,xie2021fractional,wang2021chiral,wang2021exact,bernevig2021twisted,song2021twisted,bernevig2021twistedIII,lian2021twisted,bernevig2021twistedV,xie2021twisted,kang2018symmetry,kang2019strong,po2018origin,zou2018band,po2019faithful,tarnopolsky2019origin,khalaf2019magic,bultinck2020mechanism,koshino2018maximally,parker2021field,ghiotto2021quantum,wang2020correlated,pan2020band,zang2021hartree,wang2021staggered,zang2022dynamical,wietek2022tunable,devakul2021magic} 
or similar \cite{li2021continuous,tang2020simulation,wu2018Hubbard,guerci2022chiral,li2021quantum,li2021imaging,devakul2022quantum,xie2022valley,xie2022topological,zhang2021spin} lattice constants are stacked at a small relative twist angle.
This paradigm is naturally extended to trilayer stacking and beyond, both with some layers aligned \cite{ATTLGExp,ATTLGTh,TDBLGExp,TDBLGExp2,TDBLGExp3,TDBLGTh,park2021magic} and with multiple twist angles \cite{Zhu_2020, DanieleTwistedTrilayer, Lin_2022, PRLTTLG, TTLGraphite, MOM_AFM, ArbitTTGExp}.
Recently it has also been extended to stacking at angles nearby a large commensurate twist angle \cite{dunbrack2021magic,Biao218}.
In all cases, the moir\'e pattern is obtained from layers with either the same or similar lattice constant (or a commensurate supercell).
In this paper, we lift that restriction.

We introduce moir\'e patterns made from stacking more than two layers in which no two layers separately display a moir\'e pattern.
We call these patterns ``intrinsically trilayer moir\'e" (or more generally, ``intrinsically $N$-layer moire") because, unlike twisted trilayer graphene, the moir\'e pattern disappears if any one layer is removed.
As we will explain, intrinsically trilayer moir\'e patterns cannot be viewed from the ``moir\'e of moir\'e" perspective often used to describe twisted trilayer graphene \cite{Zhu_2020}. 

Intrinsically $N$-layer moir\'e patterns have an important advantage over bilayer moir\'e patterns because they do not impose a constraint on lattice constants.
This vastly increases the space of possible material combinations.
Specifically, moir\'e patterns in bilayer systems require the constituent materials to have nearly the same lattice constant or to be nearly commensurate.
In contrast, intrinsically $N$-layer moir\'e patterns can be constructed from virtually \textit{arbitrary} combinations of materials.

In the present work, we focus on the crystal structure of intrinsically $N$-layer moir\'e heterostructures, postponing a study of electronic structure to future work.

We begin by reviewing the origin of moir\'e patterns. In Sec.~\ref{Sec:RSpMethod}, we provide an intuitive picture of how moir\'e patterns arise in real space. We explain the construction for bilayers and then offer a na\"ive generalization to multilayers. 
In Sec.~\ref{Sec:FreqMethod},
we argue that reciprocal space provides a more natural and concise characterization, from which we derive both bilayer and $N$-layer moir\'e patterns.

We then focus on multilayer heterostructures. In Sec.~\ref{Sec:TriConfigSpace}, we return to real space to resolve an apparent contradiction: the momentum-space perspective implies that periodic moir\'e patterns of more than two layers exist, but the na\"ive generalization of bilayer configuration space \cite{genkubo17,carr2018relaxation} fails to indicate these patterns, in part because the local structure is generally quasicrystalline rather than crystalline. 
Consequently, we develop a more nuanced notion of configuration space, in which some apparent degrees of freedom disappear on moir\'e wavelengths.
We discuss physical properties that are a function of this configuration space;
lattice relaxation is one example.

Finally, in Sec.~\ref{Sec:ExpProbes}, we discuss experimental probes and propose physical realizations of intrinisically $N$-layer moir\'e patterns. 

Throughout, we assume a three-, four-, or six-fold rotation symmetry shared between all layers of the moir\'e heterostructure. In the absence of this symmetry, the generic moir\'e pattern will be stripes rather than a 2D pattern.

\section{Configuration Space for Bilayers: Moir\'e Patterns in Real Space\label{Sec:RSpMethod}}

Moir\'e patterns are intuitively understood in real space as a slow modulation of the local lattice structure. 
The set of all possible local environments is known as configuration space \cite{genkubo17,carr2018relaxation}.
The configuration space approach extends beyond linear transformations of perfectly rigid crystals to include lattice relaxation effects.
However, the approach becomes subtle for heterostructures of multiple layers or different lattice constants.

In this section, we review configuration space in the simplest case of bilayers with near-identical lattices.
We then extend the formalism to bilayer systems perturbed from a commensurate stacking. Finally, we offer a ``na\"ive configuration space'' for trilayer systems, and briefly discuss how it leads to the complex patterns observed in twisted trilayer graphene. (Later, in Sec.~\ref{Sec:TriConfigSpace}, we will provide a more complete accounting of configuration space in systems with more than two layers and explain the breakdown of the na\"ive configuration space.)

\subsection{Two square lattices\label{Sec:SqBasics}}

\begin{figure}
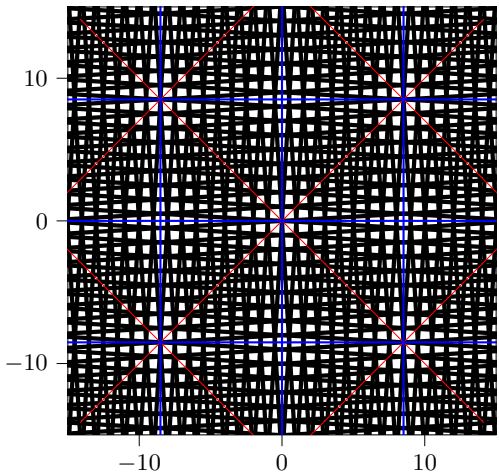

    \centering
    \include{Images_2DMoireTikZ_SqHD_SqSimple}
    \caption{A moir\'e lattice of two square layers twisted at $6.7329^\circ$. Commensurate lattice in red, moir\'e lattice in blue.}
    \label{fig:SqSimple}
\end{figure}

Consider two stacked periodic layers. There are two cases to consider: when the two layers share a common (larger) period, and when they do not. If they do share a common period, we call the structures commensurate. If they do not, we call them incommensurate.

In Fig.~\ref{fig:SqSimple}, we illustrate a small commensurate pattern formed by two square lattices at a relative twist angle of approximately $6.7^\circ$ about a square corner. This aligns the square corners of the unit cell (8,9) of one layer with (9,8) of the other, forming the commensurate superlattice outlined in red.

However, in the center of each red supercell is a location that looks very similar to the corners, where the unit cells are also aligned at the center of the square cells rather than at a vertex.
This smaller grid of locations where the square-centers are aligned defines the moir\'e lattice, outlined in blue. Thus, the visual moir\'e cell, which enjoys an approximate translation symmetry, is smaller than the commensurate unit cell, which exhibits an exact translation symmetry.
In general, the visual pattern will either be the same size or smaller than the commensurate cell 
(although for two identical square lattices, the moir\'e cell is always smaller by at least a factor of $\sqrt{2}$, regardless of twist angle.

The commensurate cell size is highly sensitive to angle and exists only on a dense subset of angles. Computing the size of a commensurate cell as a function of twist angle is analogous to determining the size of the minimal denominator of a fraction as a function of the value of that fraction, as explained in Supplement 1.

The moir\'e cell, however, varies smoothly with twist angle for small twist angles. At sufficiently large twist angles, the moir\'e cell becomes smaller than a unit cell, which indicates that the moir\'e pattern ceases to exist and no visual pattern arises.

This example shows how a moir\'e pattern arises from the two layers being stacked at different ``local relative translations'' at different positions, i.e., in the brighter regions, the lattices are stacked atom-on-atom, while in the darker regions, the lattices are stacked atom-on-void.
The moir\'e lattice is defined by the collection of points where the two layers align in either configuration.

\subsection{Local configuration space: two identical layers}

\begin{figure}
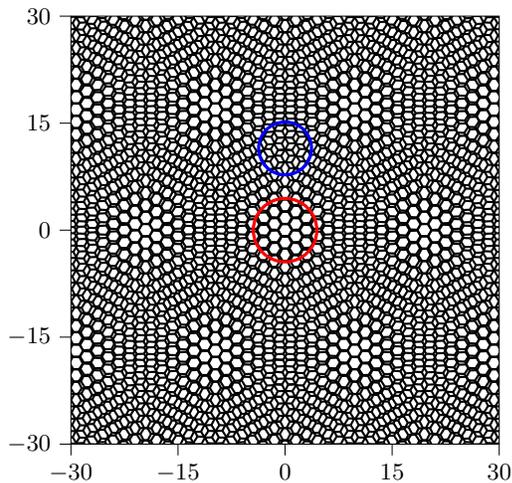

    \centering
    \include{Images_2DMoireTikZ_0DegTBLG_5DegTBLG}
    \caption{A moir\'e lattice of two hexagonal layers with unit-length interatomic distance stacked with a relative twist angle of $5^\circ$. Red and blue circles indicate an ``AA-stacked'' region where hexagons align and an ``AB-stacked'' region where they are offset, respectively.}
    \label{fig:5DegTBLG}
\end{figure}

The space of relative translations of the aligned layers defines the local configuration space. For instance, TBLG exhibits regions of AA and AB stacking, as well as intermediate regions, as illustrated in Fig.~\ref{fig:5DegTBLG}.

For two identical layers, the local configuration space is defined with respect to relative translations of the two untwisted layers, as we will now describe. 
Although the idea is intuitive in this case, developing the mathematical infrastructure carefully here will elucidate the more complicated situations we consider later.

\subsubsection{Configuration space as differences of relative coordinates\label{Sec:ConfigSpaceRelCoordsIdLayers}}

In the simplest setup where the two untwisted layers have identical lattice vectors, we define the local configuration $C(x)$ in terms of the relative coordinates $x_i$ of each layer. The relative coordinate $x_i(x)$ is a two-component vector that specifies where the position $x$ resides in the unit cell of layer $i$. Thus, $x_i$ is determined by the matrix $A_i$, whose columns are the (twisted) lattice vectors of layer $i$, as
\begin{equation}\label{eq:relcoord}
    x_i(x)=A_i^{-1}x\quad\text{mod }\I
\end{equation}
where ``mod $\I$" means ``modulo the columns of $\I$" (i.e., mod $\{(1,0),(0,1)\}$).
The local configuration is then defined as the difference between the two relative coordinates
\begin{align}
    C(x)&=x_2(x)-x_1(x)&&\text{mod }\I \label{eq:AlgebraicCSDef}\\
    &=(A_2^{-1}-A_1^{-1})x&&\text{mod }\I \label{eq:ConfigSpaceFormula}
\end{align}
While the functions $x_i$ vary on the scale of the original lattice, for a small twist or lattice mismatch, $C(x)$ varies much more slowly, and the period of $C(x)$ defines the moir\'e lattice. Therefore, the moir\'e lattice vectors are given by the columns of the matrix
\begin{equation}\label{eq:MoireLatticeVecFormula}
    A_M=(A_2^{-1}-A_1^{-1})^{-1}
\end{equation}
in the case where the inverse exists.
If the inverse does not exist, then there is not a 2D moir\'e pattern.

In the case where the two layers are identical and twisted by a relative angle $\theta$, one can simplify further by writing $A_{1,2}=R(\pm\theta/2)A$, where $R(\theta)$ is the rotation matrix. The moir\'e lattice vectors then simplify to
\begin{equation}
    A_M=\left[ R(\theta/2)-R(-\theta/2) \right]^{-1}A=\frac{1}{2\sin(\theta/2)}R\left(\frac{\pi}{2}\right)A.\label{eq:RotConfigEqn}
\end{equation}
In other words, the moir\'e lattice vectors are rotated by $\pi/2$ compared to the original lattice vectors $A$ and scaled up by a factor of $1/(2\sin(\theta/2))$.

The same formalism applies to aligned layers with a small difference in their lattice constants. For example, if $A_2=(1+\delta)A_1$, then Eq.~\eqref{eq:MoireLatticeVecFormula} can be simplified without any matrix algebra to $A_M=\frac{1+\delta}{\delta}A_1$ (neglecting the overall sign).
Generalizing to the case of two layers with a small lattice mismatch arranged with a slight twist angle yields Eq.~(1) in Ref.~\onlinecite{carr2020electronic}.

Eq.~\eqref{eq:MoireLatticeVecFormula} in this paper also allows for anisotropic lattice mismatch, as might be induced by a strain.

\subsubsection{Configuration space as a quotient of translation groups}

More abstractly, configuration space is equivalently defined as the space of nontrivial translations of the lattices before twisting, as we now explain. A combination of translations is ``trivial" if it differs from zero translation of each layer by the simultaneous translation of all layers by the same amount.

In other words: consider the two identical lattices before twisting. Denote the group of translations of each layer modulo lattice translations by $T_i$. (Note $T_i$ will be isomorphic to the torus $T^2=\R^2/\Z^2$.) Similarly denote the group of translations of the two lattices simultaneously (modulo translations that preserve the shared pre-twist lattice) as $T_{12}$. 
The space of configurations is the space of translations of each layer, modulo simultaneous translations of the two layers:
\begin{equation}
    T_\text{config}=T_1\times T_2/T_{12}.\label{eq:GeometricCSDef}
\end{equation}
This space of configurations is itself a torus.

We now relate this space to the moir\'e pattern. Suppose we transform each layer by a linear transformation $M_i$, e.g., for twist, $M_i=R(\theta_i)$. In terms of the matrices of lattice vectors before and after twisting, 
\begin{equation}\label{eq:MiDef}
    M_i=A_iA^{-1}.
\end{equation}
We now interpret this transformation as a position-dependent translation, which will give the $T_i$-coordinate in Eq.~\eqref{eq:GeometricCSDef}.

To find the translation of one layer associated with a point $x_0$ in real space, consider the map which first transforms physical space, then transforms back but centered at $x_0$. (E.g., for a twist by $\theta$, first twist about the origin by $\theta$, then twist back around $x_0$ by $-\theta$.) Conceptually, the first transformation sets up the twisted system, and the latter re-aligns the layers without further translating $x_0$. 

Algebraically, understanding that ``transform around $x_0$" can be written as ``translate $x_0$ to the origin, transform, then translate back," the translation is given by
\begin{equation}\label{eq:ConfigTranslation}
    x\rightarrow M_i^{-1}(M_ix-x_0)+x_0=x-(M_i^{-1}-\I)x_0,
\end{equation} 
which is a translation because it takes the form $x\rightarrow x-a$. This translation is then taken modulo the pre-twist lattice vectors to get the element of $T_1$.

Doing this for each layer yields the translation operators that determine a point in configuration space defined by Eq.~\eqref{eq:GeometricCSDef}. Modding out by simultaneous translations in Eq.~(\ref{eq:GeometricCSDef}) yields the relative translation difference between the two layers,
\begin{equation}\label{eq:GeomConfigSpaceFormula}
    \tilde C(x)=(M_2^{-1}-M_1^{-1})x\mod A
\end{equation}
where $A$ is the shared lattice before twisting. This is in one-to-one correspondence with the characterization of configuration space in Eq.~\eqref{eq:ConfigSpaceFormula}.
The moir\'e unit cell is given by
\begin{equation}\label{eq:GeomConfigMoireLVs}
    A_M=(M_2^{-1}-M_1^{-1})^{-1}A,
\end{equation}
which is exactly Eq.~\eqref{eq:MoireLatticeVecFormula}.
Written in this way, the moir\'e lattice is ``factored'' into one term, $M_2^{-1}-M_1^{-1}$, that depends on the transformations but not the original lattice, and another term, $A$, that depends on the lattice but not the transformations. The second term can be interpreted as the size of configuration space and the first as the rate at which the moir\'e pattern explores that space.

\subsection{Generalization to near-commensurate twisting\label{Sec:RSpCommens}}

Now instead of two identical layers, consider two layers that form a small (i.e., not moir\'e) commensurate supercell. Applying a small twist or lattice mismatch produces a moir\'e pattern. For instance, two square lattices whose side lengths differ by a factor of $\sqrt{2}$ form a commensurate supercell when arranged at a $45^\circ$ relative orientation; when twisted by an angle near $45^\circ$, they form a moir\'e pattern as illustrated in Fig.~\ref{fig:SqbySqrt2}. A second example is two identical honeycomb lattices twisted near a commensurate angle that is not a multiple of $60^\circ$, as discussed in Ref.~\onlinecite{Biao218}; near-$21.8^\circ$ TBLG is shown in Fig.~\ref{fig:218Gr}.

\begin{figure}
    \centering
    \include{Images_2DMoireTikZ_SqbySqrt2_SqbySqrt2}
    \caption{Moir\'e pattern from two square lattices with side lengths $1$ and $\sqrt{2}$ arranged with a relative twist angle of $42^\circ$.}
    \label{fig:SqbySqrt2}
\end{figure}

\begin{figure*}
    \centering
    \includegraphics[width=7in]{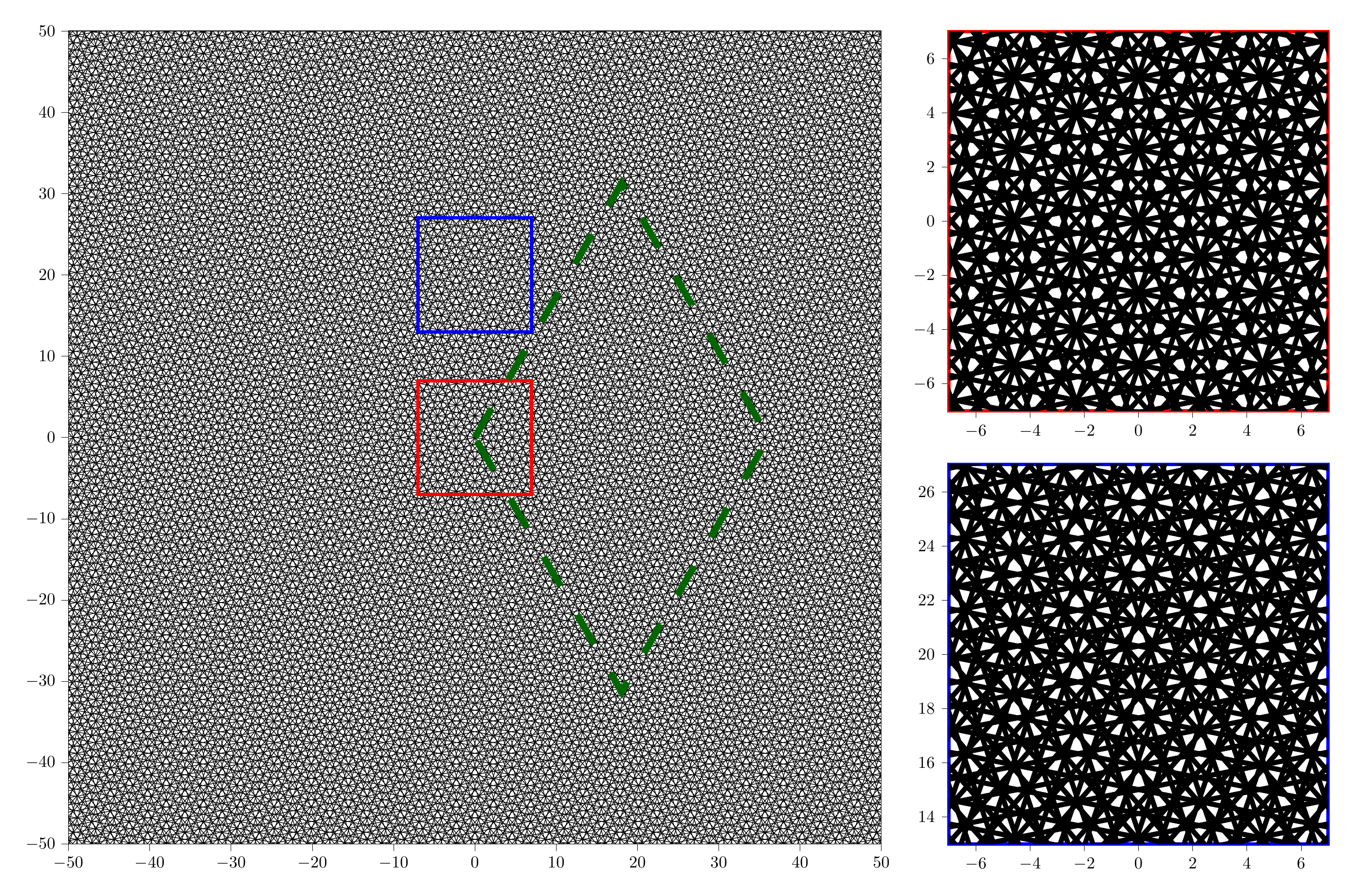}
    \caption{A moir\'e pattern formed by two unit triangular lattices arranged with a relative twist of $22.4^\circ$ ($21.8^\circ+0.6^\circ$).
    The resulting triangular moir\'e lattice has a unit cell of side length 36.1, shown in green. The moir\'e pattern is subtle, alternating between regions with individual sixfold-symmetric ``centers'' (red) and regions with triplets of ``centers'' connected in a triangle (blue). A larger picture of the moir\'e pattern is shown in Fig.~S2-1.}%\labelcref{fig:218GrUnaltered}.}
    \label{fig:218Gr}
\end{figure*}

The abstract description of configuration space described in Eq.~\eqref{eq:GeometricCSDef} extends to this case with only one minor modification: 
instead of considering the translations as acting on the lattices at zero twist, consider them at the relevant commensurate stacking. Hence, the $T_i$ are now defined modulo the individual lattices at the commensurate stacking, whereas $T_{12}$ is defined modulo the lattice vectors of the commensurate structure.

An argument for the size of the moir\'e pattern comes from Eq.~\eqref{eq:GeomConfigSpaceFormula} and the subsequent discussion. 
A linear transformation (e.g. twist) performed on a near-commensurate structure explores the configuration space at the same rate as the structure formed by performing the same transformation on a zero-degree stacked structure. However, the configuration space of the former is (perhaps counterintuitively) smaller, for reasons we now explain heuristically.

The size of configuration space in the case of two layers stacked to form a supercell can be sensibly guessed from Eq.~\eqref{eq:GeometricCSDef}. 
Let $\mathcal{A}_i$, $\mathcal{A}_C$, $\mathcal{A}_\text{cs}$ and $\mathcal{A}_M$ denote the areas of the unit cell of layer $i$, the commensurate supercell, configuration space, and the moir\'e unit cell, respectively. Replacing each translation group in Eq.~\eqref{eq:GeometricCSDef} by the area of the corresponding torus yields
\begin{equation}\label{eq:ToriToAreas}
    \mathcal{A}_\text{cs}=\frac{\mathcal{A}_1\mathcal{A}_2}{\mathcal{A}_C}.
\end{equation}
Exploiting the fact that $\mathcal{A}_i=|\det(A_i)|$ and guided by the intuition that $A$ in Eq.~\eqref{eq:GeomConfigMoireLVs} should be generalized to some ``configuration space lattice," the area of the moir\'e cell is
\begin{equation}\label{eq:CommensLatticeSize}
\mathcal{A}_M=\frac{1}{|\det(M_2^{-1}-M_1^{-1})|}\frac{\mathcal{A}_1\mathcal{A}_2}{\mathcal{A}_C}.
\end{equation}
The intuition that we should use the configuration space lattice follows from factoring Eq.~\eqref{eq:GeomConfigMoireLVs} as described in the text following that equation. (We give a rigorous description of how to find the ``configuration space lattice vectors" $A_\text{cs}$ in Appendix~\ref{Apx:NearComm} and prove that they are indeed the analogue of $A$ in Eq.~\eqref{eq:GeomConfigMoireLVs}.)

As a concrete example, consider two identical lattices twisted at an angle $\theta$ away from a commensurate stacking where the commensurate cell is a factor of $N$ larger in area than the original unit cell (for instance, in near-$21.8^\circ$ TBLG, the commensurate cell is $7$ times larger in area than the original graphene cell). The size of $T_i$ does not depend on how the layers are stacked, but $T_{12}$ will be a factor of $N$ larger in area when they are twisted $\theta$ away from the commensurate stacking compared to when the layers are stacked at an overall twist angle of $\theta$. Therefore, according to Eq.~\eqref{eq:ToriToAreas}, the configuration space, which is defined modulo $T_{12}$, would be a factor of $N$ smaller. Since the matrices $M_{1,2}$ in the denominator of Eq.~\eqref{eq:CommensLatticeSize} depend only on $\theta$ and not on the supercell or original lattice, it follows that, contrary to the most obvious intuition, for two specified 2D layers, the larger the commensurate cell, the \textit{smaller} the moir\'e pattern.

In Appendix~\ref{Apx:NearComm}, in addition to formally deriving Eq.~\eqref{eq:ToriToAreas}, the relative coordinates of heterostructures nearby a supercell configuration are derived, generalizing Eq.~\eqref{eq:GeomConfigSpaceFormula}.

\subsection{A na\"ive approach to configuration space with more than two layers\label{Sec:NaiveRSpTri}}

We now try to apply the idea of configuration space as the translation of each layer modulo overall translations to heterostructures with more than two layers. 
We call this notion ``na\"ive configuration space" (in contrast to a more nuanced notion to be given in Sec.~\ref{Sec:TriConfigSpace}). For instance, in the case of three identical layers near zero stacking, as in twisted trilayer graphene, the local configuration space is a four-dimensional torus:
\begin{equation}\label{eq:3LayerCS}
    T_\text{config}=T_1\times T_2\times T_3/T_{123}
\end{equation}
In general, the local configuration space of $N$ arbitrarily-twisted layers (with respect to a reference configuration) is a $(2N-2)$-dimensional torus:
\begin{equation}\label{eq:NLayerCS}
    T_\text{config}=\left(\prod_i T_i\right)/T_\text{all}
\end{equation}

Because this configuration space has dimension greater than two, we do not generally expect that it is fully explored. The consequence is a complex structure of overlapping moir\'e patterns (illustrated for twisted trilayer graphene in Fig.~1b of Ref.~\onlinecite{TTLG_Configurations}), and the four-dimensional space will generally be the correct parameter space for many layers twisted near a single commensurate structure of all layers (as can be seen in, e.g., Ref.~\onlinecite{DanieleTwistedTrilayer}).

As the next section will show, however, there are moir\'e patterns that arise when multilayer structures are twisted near special incommensurate configurations. 
In these cases, more care is required to define which configurations are distinct in a way that will manifest on moir\'e lengthscales: $T_\text{config}$ as written in Eq.~\eqref{eq:NLayerCS} is not correct because $T_\text{all}$ is not the correct space by which to mod out.

\section{Moir\'e in Frequency Space\label{Sec:FreqMethod}}
An alternative to defining a moir\'e pattern in real space is to define it by the appearance of low-frequency modes in momentum space. This approach is discussed at length in Ref.~\onlinecite{amidror2009theory}; here we summarize by focusing on the modes of a black-and-white image.
However, the content is much more general; see Appendix~\ref{Apx:Nonlin} for details. 

Consider a layered material as a set of transparencies placed over a light source. 
The atomic structure defines a local transmission coefficient $T_i(x)$ that specifies how much light layer $i$ lets through at point $x$. For a black-and-white image, $T_i(x)=1$ wherever the layer's image is white and $T_i(x)=0$ where it is black; this paradigm extends to grayscale images using opacities between zero and one.

By the definition of the transmission function, 
given $T_i(x)$ in each layer $i$, the resulting transmission function of the layered structure is given by:
\begin{equation}
    T(x)=\prod_i T_i(x),\label{eq:RealSpaceTransparency}
\end{equation}
which defines how the resulting multilayer pattern is formed from the patterns of the individual layers.
The moir\'e-scale physics emerges by extracting the low-frequency modes. In each periodic layer $i$, the Fourier transform is defined by:
\begin{equation}\label{eq:DefFourier}
    T_i(x)=\sum_\mathbf{n} c_{i,\mathbf{n}}\exp\left(ik_{i,\mathbf{n}}\cdot x\right)
\end{equation}
where the sum is over the reciprocal lattice vectors $k_{i,\mathbf{n}}$.

Fourier transforming Eq.~\eqref{eq:RealSpaceTransparency} yields:
\begin{equation}\label{eq:FreqSpaceTransparency}
    \hat T(k)=[\hat T_1*\hat T_2*\ldots*\hat T_N](k),
\end{equation}
where $*$ denotes the discretized convolution:
\begin{equation}\label{eq:DiscConv}
    [f*g](k)=\sum_{\mathbf{n},\mathbf{m}}c_\mathbf{n}d_\mathbf{m}\delta(k-k_\mathbf{n}-k'_\mathbf{m}),
\end{equation}
so that
\begin{equation}\label{eq:DiscConvT}
    [T_1*\ldots*T_N](k)=\sum_{\mathbf{n}_1,\ldots,\mathbf{n}_N}\left[\left(\prod_i c_{i,\mathbf{n}_i}\right)\delta(k-\sum_i k_{i,\mathbf{n}_i})\right]
\end{equation}

Therefore, a low-frequency (small-$k$) mode requires there exist a collection of modes
$\mathbf{n}_i$ so that $\sum_i k_{i,\mathbf{n}_i}\approx 0$. This sum is the moir\'e wavevector,
\begin{equation}\label{eq:moirek}
    k_M=\sum_ik_{i,\mathbf{n}_i},
\end{equation}
which in turn yields the moir\'e wavelength and orientation.

Such a collection of modes arise naturally by considering a small deformation (twist, stretch, etc.) away from a reference configuration where $\sum_i k_{i,\mathbf{n}_i}=0$ exactly.
For a bilayer system, $k_{1,\mathbf{n}}+k_{2,\mathbf{m}}=0$ is precisely a commensurability condition.
The case $\mathbf{n}=\mathbf{m}$ corresponds to the familiar near-zero-degree moir\'e pattern for nearly-identical lattices. 
On the other hand, the case $\mathbf{n}\neq\mathbf{m}$ corresponds to a near-commensurate moir\'e, which can result when the two lattices differ in size (illustrated in Fig.~\ref{fig:SqbySqrt2}) or are arranged near a commensurate angle (illustrated in Figs.~\ref{fig:218Gr} and \ref{fig:369Sq}).

\subsection{Near-commensurate example}

\begin{figure*}
    \centering
    \includegraphics[width=7in]{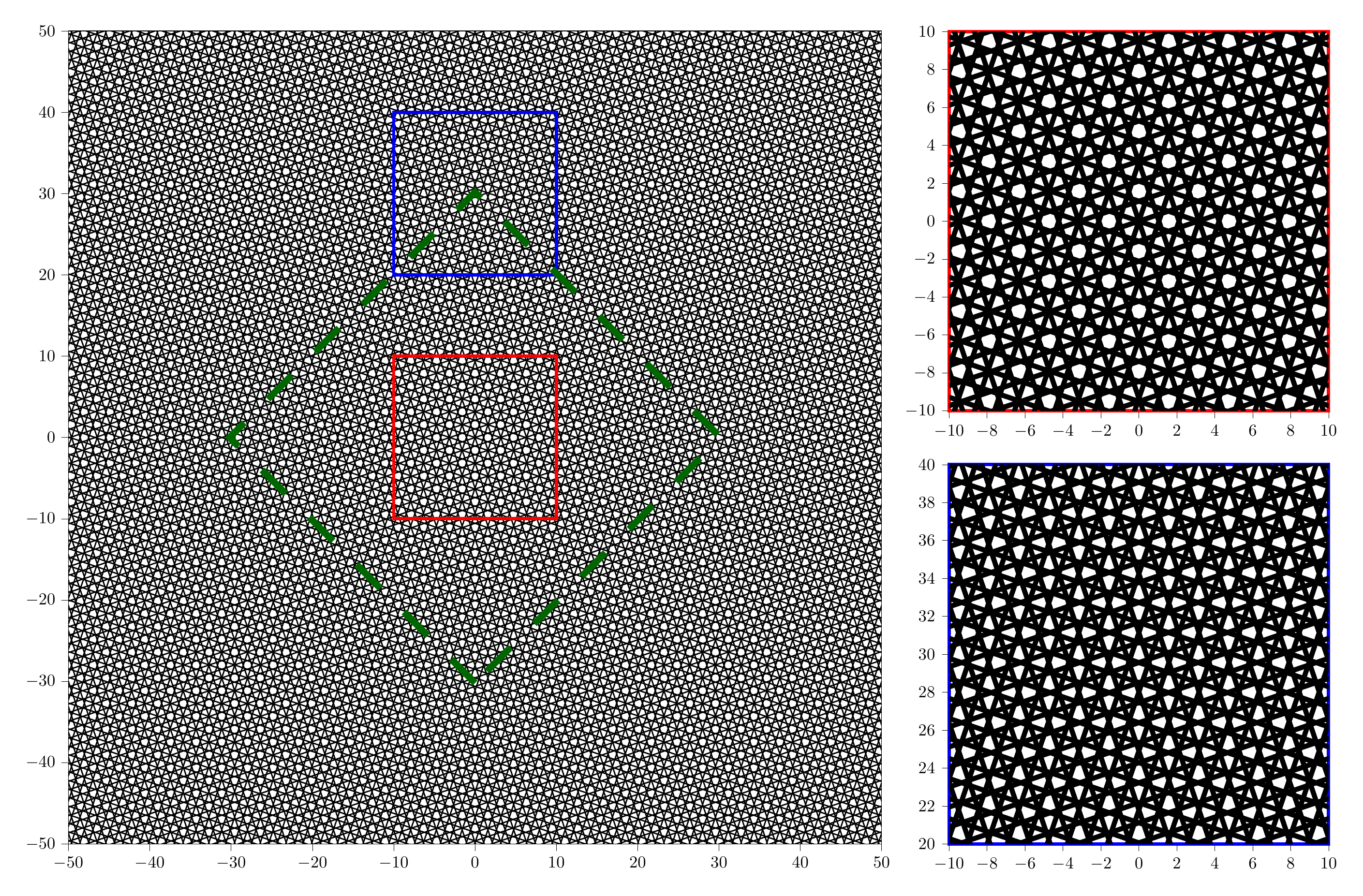}
    \caption{A moir\'e lattice formed by two unit square lattices arranged at a relative twist of $37.5^\circ$ ($36.9^\circ+0.6^\circ$), with a 42.7 side length moir\'e cell (green square). There is a resulting pattern of ``holey regions" (red square) and ``knitted regions" (blue square). A larger unannotated picture of the moir\'e pattern is presented in Fig.~S2-2.}%\labelcref{fig:369SqUnaltered}.}
    \label{fig:369Sq}
\end{figure*}

\begin{figure}
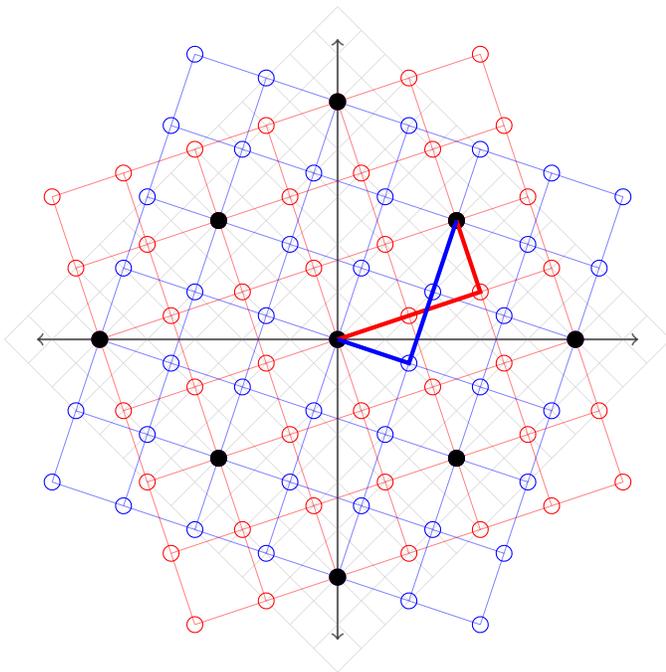

    \centering
    \include{Images_ReciprocalSpaceDiagrams_36.9RecipSpace}
    \caption{Reciprocal space of two square lattices stacked at a commensurate $36.9^\circ$ twist angle. Red(blue) open circles indicate the reciprocal lattice vectors of the top(bottom) layer; black filled circles indicate shared reciprocal lattice vectors. Thick lines shows that the (1,2) mode of the blue layer coincides with the (2,1) mode of the red layer. Light gray indicates the reciprocal commensurate lattice.}
    \label{fig:369Recip}
\end{figure}

As a concrete example, consider two square lattices arranged with a twist angle near the $36.9^\circ$ commensurate angle, as illustrated in Fig.~\ref{fig:369Sq}. The lowest Fourier modes before twisting are illustrated in Fig.~\ref{fig:369Recip}; note the (1,2) mode of one layer coincides with the (2,1) mode of the other.
The magnitude of the wave vector of these modes is $|k_{36.9}|=\sqrt{5}k_0$, where $k_0$ is the magnitude of the wave vector of the lowest mode of a single layer.

\begin{figure}
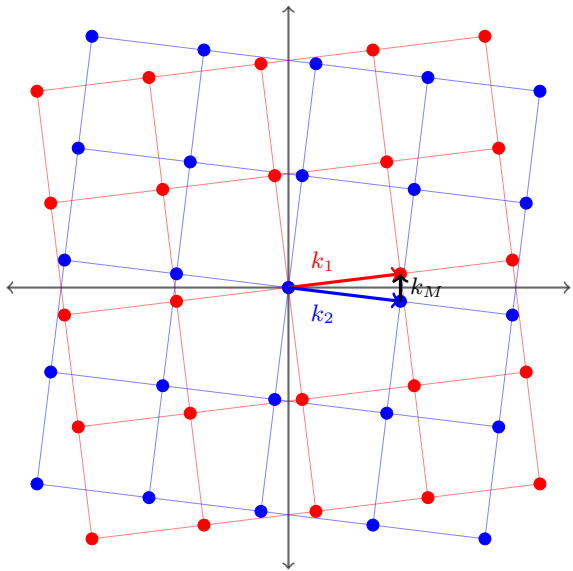

    \centering
    \include{Images_ReciprocalSpaceDiagrams_0RecipSpace}
    \caption{Lowest frequency modes of two square lattices at a small relative twist.
    Red and blue circles indicate reciprocal lattice vectors of each layer.
    The small difference between the lowest modes $k_1-k_2$ gives the moir\'e wavevector $k_M$, from which Eq.~\eqref{eq:FreqModeDiff} follows.}
    \label{fig:Near0Sq}
\end{figure}

In general, if two modes with a wave vector of magnitude $|k|$ are initially aligned before twisting, then after a relative twist by an angle $\theta$, 
the difference between the two wave vectors has magnitude
\begin{equation}\label{eq:FreqModeDiff}
    |k_M|=2\sin(\theta/2)|k|,
\end{equation}
as is seen geometrically in Fig.~\ref{fig:Near0Sq} and can be derived mathematically by taking $k_1=-R(\theta)k_2$ in Eq.~(\ref{eq:moirek}).

Accordingly, the moir\'e pattern at $36.9^\circ+\theta$ is a factor of $\sqrt{5}$ smaller in real space than the moir\'e pattern at $0^\circ+\theta$ because
\begin{equation}\begin{split}
    |k_M^{36.9+\theta}|&=2\sin(\theta/2)|k_{36.9}|\\
    &=2\sin(\theta/2)\sqrt{5}|k_0|\\
    &=\sqrt{5} |k_M^{0+\theta}|.
\end{split}\end{equation}
The same result was obtained in Sec.~\ref{Sec:RSpCommens} through more complicated arguments in real space.

The moir\'e patterns obtained from twisting near a commensurate angle, as illustrated in Figs.~\ref{fig:218Gr} and \ref{fig:369Sq}, are fainter than those for the corresponding structures near zero degrees in Figs.~\ref{fig:5DegTBLG} and \ref{fig:SqSimple}, respectively. The faint pattern occurs because the higher-frequency modes have smaller amplitudes than the lowest mode, and therefore the coefficients $c_\mathbf{n}d_\mathbf{m}$ in Eq.~\eqref{eq:DiscConv} are smaller.
(The range of visibility of different near-commensurate moir\'e patterns is also illustrated in Fig. 3.2 of Ref.~\onlinecite{amidror2009theory}.)

\subsection{Intrinsically multilayer moir\'e}

The moir\'e formalism in reciprocal space, i.e. Eq.~\eqref{eq:moirek}, also provides a requirement for a moir\'e pattern to exist in a multilayer heterostructure: there must exist a linear combination of reciprocal lattice vectors in the different layers that adds up to a vector much smaller than the reciprocal lattice vectors of the original layers. In the following, we provide a recipe for meeting this condition that is analogous to twisting near commensurate structures.

First, find a stacking arrangement of the layers such that a reciprocal lattice vector can be chosen in each layer so that the sum over the chosen reciprocal lattice vectors in all layers is zero, i.e., $\sum_i k_{i,\mathbf{n}_i} = 0$, where $k_{i,\mathbf{n_i}}$ is the chosen reciprocal lattice vector in layer $i$.
We call such a configuration \textit{singular} (following the terminology from Ref.~\onlinecite{amidror2009theory}), which is a generalization of a commensurate configuration. Note this notation differs from Ref.~\onlinecite{genkubo17}, where incommensurate is defined as non-singular in our terminology.

Once a singular configuration is identified, a small twist or stretch of each layer away from the singular configuration results in the same sum of reciprocal lattice vectors being nonzero but small. This small sum of the lattice vectors is precisely a reciprocal lattice vector of the moir\'e lattice, as defined in Eq.~\eqref{eq:moirek}.

We call a moire pattern ``intrinsically $n$-layer" if it originates from a singular configuration where no two layers are singular. In other words, an intrinsically $n$-layer moir\'e material is one whose singular configuration is a sum of reciprocal lattice vectors from all layers that add to zero, but no two vectors from that sum add to zero by themselves.
Notice this is distinct from, e.g., helically-twisted trilayer graphene~\cite{Zhu_2020, DanieleTwistedTrilayer,Lin_2022,PRLTTLG}; there the singular pattern is at zero twist angle, where \textit{any} two layers have reciprocal lattice vectors which add to zero. (Patterns where some layers are aligned, such as alternating-twisted trilayer\cite{ATTLGExp,ATTLGTh} and twisted double bilayer graphene\cite{TDBLGExp,TDBLGExp2,TDBLGExp3, TDBLGTh}, often have patterns that arise from only two misaligned sets of layers, rather than more than two; moreover, such patterns are always singular in themselves.)

\begin{figure*}
    \centering
    \includegraphics[width=7in]{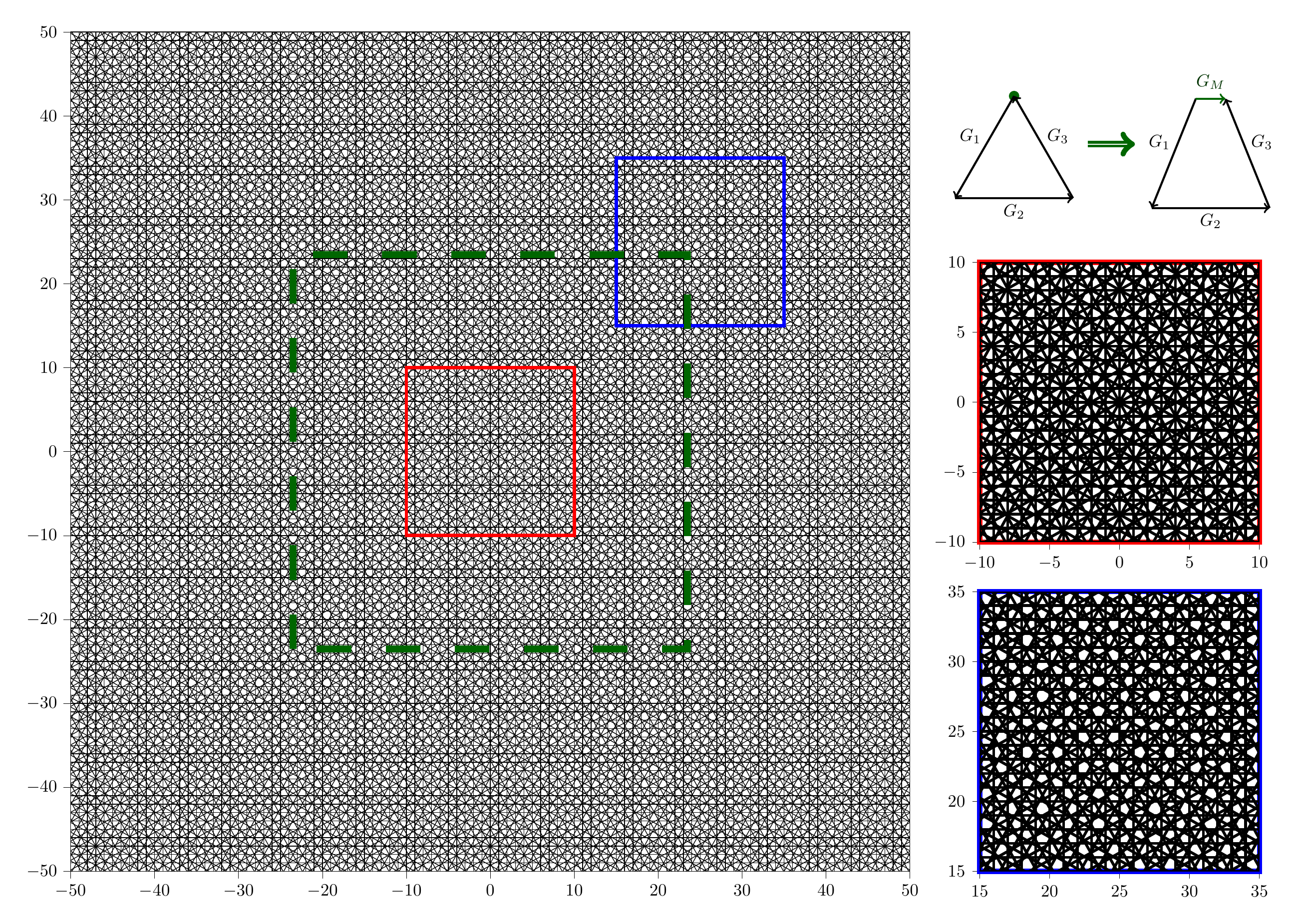}
    \caption{A moir\'e lattice of three unit square lattices at a relative twist of $119.3^\circ$, resulting in a moir\'e unit cell of side length $47$ (drawn in green). Local structures are shown at right. Top right illustrates the reciprocal lattice vectors at exactly $120^\circ$ (left) and after the $0.7^\circ$ deviation from the singular structure (right, deviation exaggerated for illustration purposes), resulting in the moir\'e reciprocal lattice vector $G_M$ shown in green. A larger unannotated picture of the moir\'e pattern is presented in Fig.~S2-3.}%\labelcref{fig:3Sq60Unaltered}.}
    \label{fig:3Sq60}
\end{figure*}

An example of an intrinsically trilayer moir\'e pattern is three square lattices twisted near $120^\circ$, illustrated in Fig.~\ref{fig:3Sq60}. The sum of the $\mathbf{n}=(1,0)$ lattice vectors from each layer vanishes, so at $120^\circ$ there is a singular structure.
Notice that this singular structure is not commensurate; in fact, it is a twelvefold-symmetric quasicrystal. In general, the singular structures will be quasicrystalline, but not necessarily with higher rotational symmetries.

\subsubsection{What is a singular structure?\label{Sec:WhatIsSingStruct}}

Since the notion of a ``singular structure" is not a standard notion of the physics literature (although it has appeared in the mathematical literature on moir\'e patterns; see Ref.~\onlinecite{amidror2009theory}), it is worth spending a moment highlighting both how it is different from a commensurate structure and how it is different from a general twist angle.

First, a multilayer system is \textit{commensurate} if the combined system has exact translation symmetries. In other words, there must exist lattice vectors $a_{1,2}$ for the multilayer system such that, for each layer $i$ with lattice vectors $a_{1,2}^{(i)}$, the vectors $a_{1,2}$ are integer linear combinations of $a_{1,2}^{(i)}$. %
As shown in Appendix~\ref{Apx:CommensPf}, this definition of commensurate is equivalent to every layer being individually commensurate with the first layer. Therefore, in an $N$ layer system with threefold or fourfold rotational symmetry, commensurability imposes $2N-2$ scalar constraints (from $N-1$ vector constraints) on the size and orientation of the lattice vectors.

By contrast, consider the singularity condition $\sum_i k_{i,\mathbf{n}_i}=0$, where $k_{i,\mathbf{n}_i}$ are each reciprocal lattice vectors of layer $i$. 
This imposes only two scalar constraints (one vector constraint) on the orientations of layers, regardless of the number of layers. For a bilayer system, the singularity condition is equivalent to commensurability, but with more than two layers, commensurability is a strictly stronger condition.

Now contrast that situation with generic twist angles. Singular structures have a property unusual among twisted systems: the average, long-distance properties of the system are sensitive to relative translations of the layers, as we now explain.

Given a system with a local property $f(x)$, the average value of that property over an area $A$ is given by $\frac{1}{|A|}\int_A f(x)d^2x$. If that area becomes very large, under appropriate convergence conditions on $f$, the average value converges to the Fourier transform of $f$ at the origin, $\hat f(0)$.

Suppose now that $f(x)$ can be written as a product of functions of each layer; e.g., for a trilayer system, $f(x)=f_1(x)f_2(x)f_3(x)$, where $f_i(x)$ is periodic with the periodicity of layer $i$. Notice the transmission function defined in Eq.~(\ref{eq:RealSpaceTransparency}) has this property.

The zeroth Fourier mode of $f$ is determined by Fourier modes $\hat{f}_i(k_i)$ of each layer such that $\sum_i k_i = 0$, as shown in Eq.~(\ref{eq:DiscConvT}). 
If the layers are not stacked in a singular structure, the only solution to $\sum_i k_i = 0$ is when $k_i = 0$ in each layer.
Therefore, the average value of $f$ in the multilayer is a product of the average values of $f$ in each individual layer;
relative translations of the layers have no impact on this zeroth Fourier mode.

By contrast, for a singular structure, there exists a nontrivial combination of Fourier modes in each layer that contribute to the average value of $f$. 
For instance, consider a trilayer system with reciprocal lattice vectors $k_i$ in each layer such that $\sum_i k_i=0$. 
Further suppose  $f_i=c_{0,i}+2c_{1,i}\cos(k_i\cdot x)$, for some coefficients $c_{0,i}$, $c_{1,i}$. From Eq.~(\ref{eq:DiscConvT}), the zeroth Fourier mode of $f$ is
\begin{equation}
\hat f(0)=c_{0,1}c_{0,2}c_{0,3}+2 c_{1,1}c_{1,2}c_{1,3}
\end{equation}
where the factor of $2$ derives from the positive and negative contributions of the cosine. (If $f_i$ had a rotation symmetry instead of being a 1D cosine, the factor of $2$ would turn into a $4$ or $6$.)
Now translating each layer $i$ by $a_i$ transforms the zeroth Fourier mode into
\begin{equation}\label{eq:ZerothFourierPhase}
\hat f(0)=c_{0,1}c_{0,2}c_{0,3}+2c_{1,1}c_{1,2}c_{1,3}\cos(\sum k_i\cdot a_i),
\end{equation}
which is different for generic choices of $a_i$.

Thus, the physical consequence of a singular structure is that local properties of the multilayer are sensitive to relative translations.
This is also true for commensurate structures, but is not true for a general non-singular or non-commensurate stacking.
However, notice that for a fixed set of $k_i$, Eq.~\eqref{eq:ZerothFourierPhase} is invariant under the special set of translations $a_i$ which satisfy $\sum k_i a_i = 0$. These special translations will be important in developing our notion of configuration space for multilayer systems in Sec.~\ref{Sec:TriConfigSpace}.

As discussed in Sec.~\ref{Sec:FreqMethod}, the condition that the physical quantity of interest is a product of properties in each layer, i.e., $f=f_1f_2f_3$ for a trilayer system, simplifies the discussion, but can also be relaxed significantly. The more general description is given in Appendix~\ref{Apx:Nonlin}.

\subsubsection{Labelling singular structures\label{Sec:SingStrucLabels}}
We now provide a convenient labelling schema for singular structures. Since a singular structure is specified by a combination of reciprocal lattice vectors that adds up to zero, it can be conveniently labelled by the integer indices of the reciprocal lattice vectors.

Let $b_{i,1}$ and $b_{i,2}$ be the basis of reciprocal lattice vectors in layer $i$. Then a singular structure will be specified by a set of $n_{i,j}$ that satisfy the singularity condition \begin{equation}\label{eq:SingularityCondition}
    \sum_{i,j} n_{i,j}b_{i,j}=0.
\end{equation}
For a trilayer system, the singular structure given by $n_{i,j}$ is labelled as $(n_{1,1},n_{1,2};n_{2,1},n_{2,2};n_{3,1},n_{3,2})$. This description can be generalized to any number of layers, including bilayers.
Note that the labelling depends on the choice of reciprocal lattice vectors; thus, a set of $n_{i,j}$ combined with knowledge of the reciprocal lattice vectors in each layer determines the singular structure.

The $n_{i,j}$ for an $N$-layer system naturally live in $\Z^{2N}$. 
The singularity condition in Eq.~\eqref{eq:SingularityCondition} defines a 1D sublattice in this space. Assuming rotational symmetry, one choice of $n_{i,j}$ yields another linearly-independent $n_{i,j}$ after rotation.
Thus, combined there is a 2D sublattice in $\mathbb{Z}^{2N}$ satisfying the singularity condition. 
It is also possible for the sublattice to have a higher  even dimension, as we will show for trilayer graphene in Sec.~\ref{Sec:DoublySingular}.
Regardless of dimension, we call the $n_{i,j}$ that satisfy the singularity condition the \textit{zero mode lattice}, because they correspond to combinations of Fourier modes in each layer that contribute to the $k=0$ Fourier mode of the singular structure.
Under the assumption that the sublattice is 2D and that the degree of rotational symmetry is known, each singular structure can be labelled by a single set of $n_{i,j}$ that defines one of the basis vectors of the zero mode lattice; the other basis vector follows from rotational symmetry.  %

As a few concrete examples: the standard near-zero moir\'e pattern of two layers is the $(1,0;-1,0)$ moir\'e pattern because $b_{1,1}-b_{2,1}=0$. The near-$21.8^\circ$ structure shown in Fig.~\ref{fig:218Gr} and the near-$36.9^\circ$ structure in Fig.~\ref{fig:369Sq} are both $(1,2;-2,-1)$ moir\'e patterns because $b_{1,1}+2b_{1,2}-2b_{2,1}-b_{2,2}=0$ in both cases, despite their different rotational symmetry. Finally, the intrinsically trilayer pattern illustrated in Fig.~\ref{fig:3Sq60} would be the $(1,0;1,0;1,0)$ moir\'e, assuming the first basis vector of the three layers are chosen 120 degrees apart.

\subsubsection{Degeneracy of singular structures\label{Sec:Degeneracy}}
We now consider how singular structures arise in the manifold of possible twists and lattice mismatches between the layers, which we call \textit{deformation space}. (More generally, we could also include strains that break rotational symmetries in our deformations; we call this generalization \textit{anisotropic deformation space}. However, since such deformations can result in 1D instead of 2D moir\'e patterns, we neglect such transformations here and simplify our discussion by referring to our space of isotropic deformations by the shorter term.)

Commensurate structures of bilayer systems are special among singular structures because they are zero-dimensional manifolds in deformation space: no small deformation of a bilayer singular structure yields the same singular structure. For instance, in the simple case of aligned layers (corresponding to the $(1,0;-1,0)$ commensurate structure), no combination of small relative mismatch or twist of the two layers will yield another $(1,0;-1,0)$ commensurate structure.

This is not, however, the case for singular structures with more than two layers. With $N$ layers there are $2N-2$ possible isotropic deformations (twists and isotropic strains) of the layers relative to each other: each layer beyond the first adds two additional parameters (namely, strain and mismatch with respect to the first layer). The singular structure then adds two constraints (Eq.~\eqref{eq:SingularityCondition} and its rotated counterpart) on this deformation space, meaning that it forms a $(2N-4)$-dimensional manifold in this space of deformations.

\begin{figure}
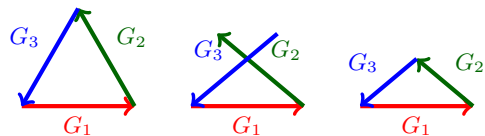

    \centering
    \include{Images_Other_TriangleSingularStructures}
    \caption{Starting from a particular singular structure, a small twist away combined with a corresponding strain results in another singular structure. These transformations yield a manifold of singular structures rather than an isolated point, as occurs for bilayers.}
    \label{fig:TriSingStruct}
\end{figure}

Intuitively, this is because there is a continuum of ways to change the sides of the triangle that keep it a triangle. 
For example, given a triangle formed by reciprocal lattice vectors, one can deform two of the lattices by a combination of twists and (isotropic) strains while leaving the third fixed and still have a triangle, as illustrated in Fig.~\ref{fig:TriSingStruct}.
In contrast, the only way to deform the layers and preserve a singular digon formed by the reciprocal lattice vectors of a bilayer is to perform an overall twist or isotropic stretch of both layers simultaneously. 

These singularity-preserving deformations are at the crux of understanding what the na\"ive configuration space description in Sec.~\ref{Sec:NaiveRSpTri} fails to see about intrinsically trilayer moir\'e patterns,  namely, why the effective parameter space seems to be periodically spanned by the two dimensional moir\'e pattern even though the na\"ive parameter space is four-dimensional. The connection between these pictures will be explained in Sec.~\ref{Sec:SingStrucDegenConfigSpaceConnection}.

\subsection{The doubly-singular structure of twisted trilayer graphene\label{Sec:DoublySingular}}
We now examine twisted trilayer graphene from the perspective of singular structures. Twisted trilayer graphene arises at the intersection of \textit{two} singular structures: the $(1,0;-1,0;0,0)$ singular structure and the $(0,0;1,0;-1,0)$ singular structure. In this sense, it is ``doubly-singular"; therefore, with four singularity constraints instead of the two considered in the previous section, the combination of singular structures is zero-dimensional, not 2D like the intrinsically trilayer pattern (the dimension is $2N-6$ instead of $2N-4$, where $N=3$ for three layers).

Twisting relative to the singular structure in this case be understood as generating multiple moir\'e patterns simultaneously.
Without a fine-tuned combination of twist and mismatch, the overlapping structure of the multiple moir\'e patterns complicated quasiperiodic patterns, as illustrated in Refs.~\onlinecite{TTLG_Configurations, ArbitTTGExp}.

In the special case where the twist angles of the first and third layers are equal and opposite, however, something special happens: at $\frac{1}{\theta^2}$ length scales, a single regular moir\'e pattern is observed.
This pattern is referred to as a ``moir\'e of moir\'e," since it arises from a moir\'e pattern induced by the two competing $\frac{1}{\theta}$-scale moir\'e patterns.

This $\frac{1}{\theta^2}$-order pattern can be understood as the pattern arising from the $(1,0;-2,0;1,0)$ singular structure. 
Specifically, defining $k_0$ to be a smallest reciprocal lattice vector of graphene, 
the trilayer structure where the first and third layers are twisted a small amount in opposite directions away from the middle layer can be described by
$k_1=R(\theta)k_0$, $k_2=-2k_0$, and $k_3=R(-\theta)k_0$.
Per Eq.~\eqref{eq:moirek}, the moir\'e wave vector is given by
\begin{equation}
    k_M=[R(\theta)+R(-\theta)-2\I]k_0=2(\cos(\theta)-1)\mathbb{I} k_0,
\end{equation}
which is of order $\theta^2$ for small $\theta$. Hence, the moir\'e wavelength is of order $\frac{1}{\theta^2}$.

\begin{figure}
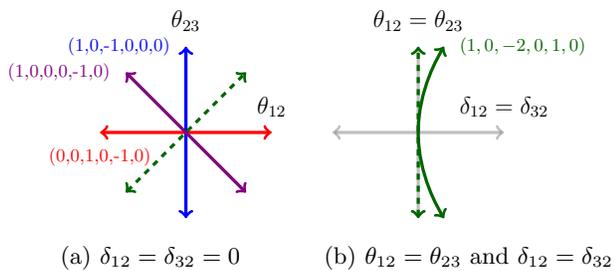

    \centering
    \begin{subfigure}{0.23\textwidth}
        \resizebox{\linewidth}{!}{\include{Images_Other_TTLGSingStruct}}
        \caption{$\delta_{12}=\delta_{32}=0$
        }
    \end{subfigure}
    \begin{subfigure}{0.23\textwidth}
        \resizebox{\linewidth}{!}{\include{Images_Other_TTLG121SingStruct}}
        \caption{$\theta_{12}=\theta_{23}$ and $\delta_{12}=\delta_{32}$}
    \end{subfigure}
    \caption{Several singular structures of TTLG along two specific slices of the four-dimensional parameter space $(\theta_{12},\theta_{23},\delta_{12},\delta_{32})$ indicated by solid colored lines. The dashed green line represents the constraint of helically-twisted trilayer graphene. The bilayer singular structures shown in the left figure deviate from helically-twisted trilayer graphene to order $\theta$, but the trilayer singular structure shown in the right figure only deviates to order $\theta^2$. Hence, the green singular structure produces a moir\'e pattern at $1/\theta^2$-scale, whereas the bilayer singular structures plotted in blue/red/purple produce (competing) moir\'e pattern at $1/\theta$ scale.}
    \label{fig:TTLGSingStructs}
\end{figure}

Moreover, since the order-$\theta^2$ deviation is only from this particular singular structure, and not from the ``doubly-singular'' structure, it exhibits a single 2D moir\'e pattern rather than complex overlapping structures. The relevant singular structures are illustrated in Fig.~\ref{fig:TTLGSingStructs}.

\section{Configuration Space of Intrinsically Trilayer Moir\'e Patterns\label{Sec:TriConfigSpace}}

There is an apparent contradiction between the na\"ive configuration space described in Sec.~\ref{Sec:NaiveRSpTri}, which indicates that trilayers have complex moir\'e patterns that cannot possibly fit on a lattice, and the intrinsically trilayer moir\'e patterns presented in Sec.~\ref{Sec:FreqMethod}, which very clearly do so. We seek to resolve this contradiction by a more nuanced description of the configuration space.

The missing ingredient from the na\"ive configuration space given in Eq.~\eqref{eq:NLayerCS} is a collection of ``nontrivial trivial transformations,'' which are nontrivial in that they do not correspond to overall translations, but trivial in that they do not change the local moir\'e structure. The correct configuration space of the moir\'e pattern is the set of translations of each layer modulo overall translations (i.e., simultaneous translations of all layers by the same amount) \textit{and} these new transformations.

We now describe how to find these additional transformations. We do so in a way that naturally derives not only the dimensionality of the true configuration space, but also explains why it is toroidal.

The intuition of the argument derives from the characterization of singular structures provided in Sec.~\ref{Sec:WhatIsSingStruct}: singular structures are those structures for which certain relative translations of the layers change the average value of local quantities by providing phases between different contributions to the zeroth Fourier mode of the quantity of interest, as in Eq.~\eqref{eq:ZerothFourierPhase}. A moir\'e heterostructure can be viewed as resulting from these different possible phases: different regions in the moir\'e heterostructure correspond to different relative translations of the singular structure.

The nontrivial trivial transformations we seek to find derive from the converse of that identification: any relative translation which does \textit{not} result in a phase will make no impact on average properties. Such relative translations that do not result in phases, therefore, are precisely the nontrivial trivial transformations.

We find the nontrivial trivial transformations formally using in the frequency picture described in Sec.~\ref{Sec:FreqMethod}.
For simplicity, we take as a concrete example the $(1,0;1,0;1,0)$-moir\'e on the square lattice (illustrated in Fig.~\ref{fig:3Sq60}). The Fourier modes are indexed by $\Z^6$, but the moir\'e modes arise from the zero mode lattice described in Sec.~\ref{Sec:SingStrucLabels}. In this specific case, the zero mode lattice is spanned by the vectors $(1,0,1,0,1,0)$ and $(0,1,0,1,0,1)$, which we call $n^{(1)}$ and $n^{(2)}$ (each of which also have indices, $n_{i,j}^{(1,2)}$).

A translation of layer $i$ by  $\mathbf{a}_i$ (not necessarily a lattice vector) will multiply the Fourier mode with indices $n_{i,j}$ by a phase $\exp(\sum_{i,j}n_{i,j}\mathbf{b}_{i,j}\cdot\mathbf{a}_i)$, which follows from the discrete Fourier transform in Eq.~\eqref{eq:DiscConvT}.
For the relative translations which preserve the moir\'e lattice, this phase vanishes when evaluated on the zero mode lattice.

Clearly, translating each layer by the same amount, $\mathbf{a}_i=\mathbf{a}$, results in this phase vanishing on the zero-mode lattice, where $\sum n_{i,j}^{(k)}\mathbf{b}_{i,j}=0$ for both $k$.
This imposes two constraints on the six-dimensional space.

The additional constraints are found by setting $\mathbf{a}_1=0$, at which point the constraint is $\mathbf{b}_{2,i}\cdot \mathbf{a}_2=-\mathbf{b}_{3,i}\cdot \mathbf{a}_3$; the simplest two basis solutions are $\{\mathbf{a}_2=\mathbf{b}_{3,1},\mathbf{a}_3=-\mathbf{b}_{2,1}\}$ and $\{\mathbf{a}_2=\mathbf{b}_{3,2},\mathbf{a}_3=-\mathbf{b}_{2,2}\}$. 
These extra translations are most of the ``nontrivial trivial transformations" we were searching for, and suffice to reduce the dimensionality of the configuration space from four to two. %
Note that this two-dimensional space is periodic, i.e., a torus rather than a plane, because the sum $\sum_{i,j}n_{i,j}\mathbf{b}_{i,j}\cdot\mathbf{a}_i$ need not vanish identically for the phase to vanish; instead, it can be a multiple of $2\pi$. 
This periodicity ensures that the final phase space is indeed a torus.

Therefore, our final and most general characterization of the phase space is as the collection of relative translations of the layers modulo those which act trivially on the zero mode lattice (i.e., on the combinations of modes that contribute to the zero mode in the singular structure).

Note that in multiply-singular structures, such as TTLG, the moir\'e-generating lattice is greater than two-dimensional. Therefore, there is at least a four-dimensional manifold defining the configuration space.
Consequently, one cannot regard this configuration space as being periodically fully explored in real space.
This explains the difference between the complex patterns in TTLG and the periodic moir\'e in intrinsically trilayer systems. %

\subsection{Configuration space and lattice relaxation \label{sec:LatRelax}}

To illustrate the usefulness of configuration space, we consider lattice relaxation in intrinsically trilayer moir\'e systems. 
Lattice relaxation is usually computed by taking an average energy density of any particular stacking configuration, then enlarging regions of low-energy stacking while shrinking regions of high-energy stackings \cite{cazeaux2018energy}.

We claim that \textit{the average energy density of a singular structure on long wavelengths does not change under a nontrivial trivial transformation}. That is to say, structures in the na\"ive configuration space (Sec.~\ref{Sec:NaiveRSpTri}) that differ by a nontrivial trivial transformation have the same energy density.

We now justify this claim.
Consider the energy density of a singular structure, $\rho(x)$, and consider the energy density over some large region of radius $R$, $\frac{1}{\pi R^2}\int_{|x|<R} \rho(x)d^2x$. As $R\rightarrow\infty$, this is precisely the zero-frequency mode of the Fourier transform of energy density, $\hat\rho(k=0)$.
Since the nontrivial trivial transformations preserve the zero mode lattice, they preserve any observable that is only dependent on that mode.
In particular, they do not change $\hat\rho(k=0)$. Therefore, the average density is only dependent on the reduced configuration space, not the higher-dimensional na\"ive configuration space.

In a moir\'e heterostructure, this implies that long-wavelength lattice relaxations arise on the moir\'e scale. A particular point $x$ on the moir\'e pattern specifies a specific stacking of the singular structure; let $\rho_x$ denote the local average energy density for that singular structure. On length scales much longer than atomic lengthscales but much shorter than the moir\'e lengthscale, we can approximate the local energy density by $\hat\rho_x(k=0)$, i.e., the average energy density of the singular structure formed at the point $x$. This is a moir\'e-periodic function of $x$, and since long-wavelength lattice relaxation can be extracted from the energy density, long-wavelength relaxations are periodic on the moir\'e lengthscale.

\subsection{Relation to singular structure degeneracy\label{Sec:SingStrucDegenConfigSpaceConnection}}

We now connect the set of nontrivial trivial transformations to the singular manifold in deformation space (discussed in Sec.~\ref{Sec:Degeneracy}). In short, while we have so far been considering a structure as a deformation from a particular singular \textit{point}, it is more accurate to consider a structure as a deviation from the \textit{manifold} of singular structures generated by including stretch as well as twist. The nontrivial trivial transformations correspond to moving along this manifold.

To understand the relationship, we begin by extending the discussion surrounding Eq.~\eqref{eq:ConfigTranslation}. Take a singular structure and transform each layer $i$ by a matrix $M_i$. Then consider the matrix (written here for a trilayer in terms of $2\times 2$ blocks)
\begin{equation}\label{eq:MOverallDef}
    M=\begin{bmatrix}M_1^{-1}-\I\\M_2^{-1}-\I\\M_3^{-1}-\I\end{bmatrix}.
\end{equation}
Similar to how in a small-angle twisted bilayer, each point in real space can be viewed as a specific untwisted stacking of the two layers, each point in a near-singular trilayer heterostructure can be viewed as a specific singular stacking of the three layers. 
The matrix $M$ maps a point in real space to the relative translation of layers required to transform the stacking at the origin to the stacking at that particular point.

The specific matrices $M'$ which map $\R^2$ to nontrivial trivial transformations (or overall translations) are precisely those which correspond to deformations $M_i$ that \textit{preserve the singular structure} of our setup (i.e., those that move along the degenerate manifold of singular structures, rather than perturbing off of it).
This makes sense because the low-frequency moir\'e modes arise from deviations from the singular structure; therefore, moving along the singular structure manifold does not yield moir\'e.

\begin{figure}
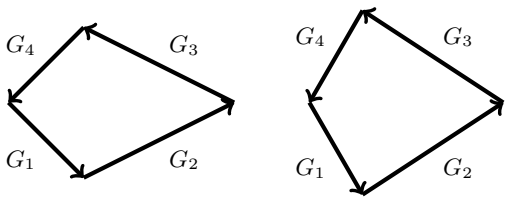

    \centering
    \include{Images_Other_TwoKiteSingularStructures}
    \caption{Different quadrilaterals can be formed with the same side lengths.
    Consequently, a stacked four-layer system can have multiple singular configurations with different large twist angles, i.e., there are different twist angles such that $\sum_i G_i = 0$, where $G_i$ is a reciprocal lattice vector in each layer. Moir\'e lattices formed by twisting slightly away from these configurations exhibit the same physics on the moir\'e length scale, provided the small twists are chosen to give the same moir\'e lattice vectors.}
    \label{fig:kitechoices}
\end{figure}

This identification has important implications for twisted multilayer moir\'e systems beyond trilayers. In a trilayer system, given a reciprocal lattice vector from each layer, there is a unique way to stack the layers (i.e., a unique set of twist angles) that results in a singular structure, provided such a structure is possible. This results from the fact that given three sides of a triangle, the interior angles of the triangle are determined. However, for four or more sides, the side lengths do not uniquely specify the interior angles, as shown in Fig.~\ref{fig:kitechoices}. Consequently, in a heterostructure with four or more layers, there are multiple twist angles that result in a singular structure.

The same moir\'e lattice can be formed by twisting away from either of these configurations.  Since equivalent points on the two moir\'e patterns differ only by a nontrivial trivial transformation, their moir\'e-scale physics is identical. This is elaborated in Appendix~\ref{Apx:Nonlin}.

A similar phenomenon occurs in a trilayer system if slight strain is included, i.e., for a given three layers, multiple singular configurations are possible if the layers can be isotropically strained in addition to being twisted.
It may be possible to make use of the choice in reference configuration for theoretical insight or computational advantage, as discussed briefly in Appendix~\ref{Apx:OptSingStruct}.

\section{Detection of Intrinsically Trilayer Moir\'e\label{Sec:ExpProbes}}
We now describe how to measure intrinsically multilayer patterns experimentally. 
A measurement that sees the moir\'e pattern must probe each layer: in an intrinsically multilayer moir\'e structure, no subset of layers alone will exhibit a moir\'e pattern, unlike a trilayer moir\'e of moir\'e structure.

\subsection{Structural probes}
One standard way to detect moir\'e patterns in bilayer systems is to use STM. 
However, a surface probe like STM primarily probes the top layer of a heterostructure.
This effectively probes the moir\'e pattern in a bilayer system because the top layer reconstructs on the moir\'e scale.
However, such a reconstruction may be weak in a multilayer system due to the large twist angles and multiple layers.
Thus, we expect STM to be less effective at probing intrinsically trilayer moir\'e patterns than at probing bilayer patterns.

In contrast, we expect TEM -- which has already been used to detect moir\'e patterns in bilayer systems \cite{TEMMoire} -- to be an ideal probe because it passes through all the layers.
TEM does not require lattice relaxation to see the moir\'e effect: the pattern that results from diffraction from each layer sequentially reproduces the sums of reciprocal lattice vectors discussed in Sec.~\ref{Sec:FreqMethod} (see Eq.~\eqref{eq:moirek}).
Therefore, intrinsically-trilayer structures will produce satellite peaks.

\subsection{Transport: engineering flat bands using intrinsically trilayer moir\'e}

Transport probes of intrinsically multi-layer moir\'e patterns depend strongly on the electronic structure of the underlying materials.
A full study of engineering electronic structures from intrinsically trilayer moir\'e is beyond the scope of this work.
Instead, we propose a few promising platforms.

\subsubsection{Large-angle TBLG with a potential}
As a first setup, consider twisted bilayer graphene at a large angle. Unlike small-angle TBLG, if the $K$ points of the two layers are significantly separated in momentum space after twisting, then interlayer hopping will couple only to high-energy states.

\begin{figure}
    \centering
    \includegraphics[width=3in]{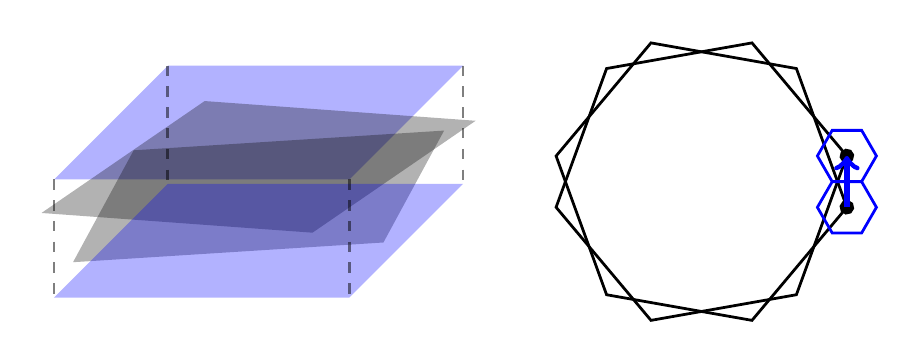}
    \caption{Two layers of graphene (black) arranged with a large twist angle generate a moir\'e pattern with the additional outer (blue) layers on the exterior, which are aligned with each other.   
    Left: Physical configuration of the layers. Right: The large black hexagons and small blue hexagons indicate the BZ of graphene and the outer layers, respectively. The reciprocal lattice vector of the outer layers couples the $K$ points of the graphene layers, effectively compensating for the large twist.}
    \label{fig:LargeAngleTBG}
\end{figure}

This obstacle is overcome by sandwiching the large-angle TBLG between two copies of a third insulating layer chosen so that its reciprocal lattice vectors (almost) perfectly compensate the momentum difference between the $K$ points of the two graphene layers. This setup is shown in Fig.~\ref{fig:LargeAngleTBG}.

If an electron feels a potential from the insulating layer as it hops from one graphene layer to the other, then it can hop from $K$ in one layer to (nearly) $K$ of the other, mimicking the process in TBLG. However, the resulting system is slightly different from magic angle TBLG because the Dirac cones are rotated with respect to each other. (The relative rotation of the Dirac cones in magic angle TBLG is small enough to be ignored.)

A different large angle moir\'e bilayer graphene structure was studied in Ref.~\onlinecite{Biao218}.
There it was found that with one tuning parameter, a ``hyper-magic manifold" with many flat bands and a kagome-like band structure emerges. 
In that paper, however, the authors were limited by needing to be near a commensurate structure. %
Our proposal described above avoids that limitation, at the cost of requiring a suitable third material.

\subsubsection{Two potentials imposed on a single layer}

Consider a layer of graphene sandwiched between two identical insulators.
If the insulating layers are arranged at a small relative angle, then they will impose a superlattice potential on graphene, whose size is determined by the moir\'e scale of the two layers. The effect of a superlattice potential on graphene has been extensively studied \cite{DeanSuperlattice,GrSL1,GrSL2,GrSL3,GrSL4,GrSL5,GrSL6,GrSL7,GrSL8,GrSL9}. 
Notably, an artificially imposed potential has been shown to produce satellite cones on a single layer of graphene \cite{DeanSuperlattice} and is predicted to produce topological flat bands in bilayer graphene \cite{SayedBLGSL}. A superlattice potential on the surface of a topological insulator may also induce correlated topological phases \cite{TwistedTISuperlattice,TISuperlatticeMerons,FuTISuperlattice}.

Alternately, the outer layers can be arranged to form an intrinsically trilayer moir\'e pattern with the center layer.
This set-up should yield the same band structure as the previous proposal, but with two physical differences.
First, this structure's existence is now dependent on the orientation of the lattice in the center layer. This dependence on the center layer enables more tunability but requires additional control. Second, lattice relaxation effects between the two insulating layers are likely to be very small since they are not arranged at a small angle. Theoretically, this lack of relaxation indicates that a rigid rotation approximation is generally more accurate than a 1D network limit (studied in, e.g., Refs.~\onlinecite{Huang_2018,KimTBGDefectNetwork,Networks1D}). 

\begin{figure}
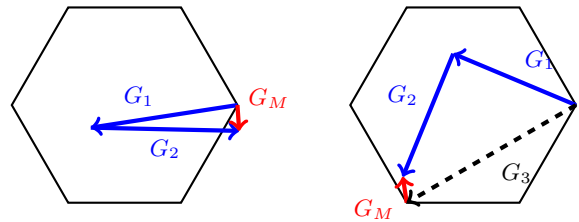

    \centering
    \include{Images_ExpSetups_TwoPotentialGraphene}
    \caption{Inducing two periodic potentials (reciprocal lattice vectors in blue) on a layer of graphene (BZ in black) produces a moir\'e superlattice (reciprocal lattice vector in red). Left: sandwiching graphene between two nearly-aligned layers produces an effective superlattice potential. The alignment of the graphene layer is unimportant. Right: if the two other layers are twisted at a large angle so their reciprocal lattice vector adds to one of graphene, intrinsically-trilayer moir\'e can arise.}
    \label{fig:TwoPotGraphene}
\end{figure}

The two setups are compared in Fig.~\ref{fig:TwoPotGraphene}.

\section{Conclusions}
We have presented a new kind of moir\'e structure, ``intrinsically trilayer moir\'e," which results from twisting multilayers near certain special ``singular structures.'' 
The local structure of such systems is quasicrystalline, but this quasicrystalline structure is periodically modulated on long (moir\'e) lengthscales.

We characterized the local configuration space of such systems, and showed that previous description of configuration space for bilayer systems \cite{genkubo17,carr2018relaxation} is insufficient to provide a real-space intuition for why these patterns arise.
Our new notion of configuration space is useful to determine lattice relaxation effects.
It also explains the $\frac{1}{\theta^2}$ moir\'e pattern in helically-twisted trilayer graphene \cite{Zhu_2020}.

Finally, we connected these abstract patterns to their material realizations. We described how to observe intrinsically multi-layer moir\'e structures experimentally, contrasting STM and TEM probes' suitability for this purpose. We also proposed a few promising material realizations that may give rise to flat bands via either interlayer or intralayer hopping terms. Other possible future directions would be to examine higher-order interlayer hopping processes or layers that are individually strongly interacting.

The systems we propose thus far are the simplest cases, and don't take advantage of the most potent aspect of these patterns: the ability to engineer moir\'e heterostructures \textit{without regard for lattice constant}. Intrinsically trilayer moir\'e can be made from materials with \textit{any} lattice constant combination, and therefore enables engineering moir\'e heterostructures with material combinations not previously imaginable, including those where the individual layers have vastly different physics.

\section{Acknowledgments}
We thank Cory Dean, Philip Kim, Abhay Pasupathy, and Ziyan Zhu for useful conversations. 
This material is based upon work supported by the National Science Foundation under the Columbia MRSEC on Precision-Assembled Quantum Materials (PAQM), Grant No. DMR-2011738.
This work was performed in part at the Aspen Center for Physics, which is supported by National Science Foundation grant PHY-1607611.
J.C. acknowledges the support of the Flatiron Institute, a division of the Simons Foundation, and the Alfred P. Sloan Foundation through a Sloan Research Fellowship.

\bibliographystyle{apsrev4-2}
\bibliography{main}

\appendix

\section{Genericity of moir\'e frequencies: beyond black \& white images\label{Apx:Nonlin}}

In this section, we explain why moir\'e frequencies are generic across a wide class of observables and not specific to those that obey Eq.~\eqref{eq:RealSpaceTransparency}. 

We first demonstrate that the moir\'e periodicities are generic (i.e., not specific to products only) when composed from quantities periodic in each layer. The amplitude of the resulting Fourier modes will change, but the frequencies themselves will remain the same. These arguments generally follow those provided in in Ch. 2.2-2.3 of Ref.~\onlinecite{amidror2009theory}.

Then, we examine functions which are periodic on the atomic scale only up to a phase, i.e., which are composed of quantum operators located at points other than $\Gamma$ in the Brillouin zone. In this case, the same argument applies with slight adaptation. This argument also shows which combinations of points in the Brillouin zone provide operators that vary on the moir\'e lengthscale.

\subsection{Generality of moir\'e frequencies for periodic functions}
We begin with the case where the observable of interest is a combination of periodic functions of the individual layers, but where the combination rule is not multiplicative as in Eq.~\eqref{eq:RealSpaceTransparency}. For simplicity, we will focus on bilayers, but the argument easily extends to multilayers.

Assume our property of interest is described by $f(g_1(x),g_2(x))$, where $g_i(x)$ is a function with the periodicity of the $i$th layer and $f$ is any function. In the case of black-and-white images discussed in the main text, $g(x)$ is the transmission function and $f(x,y)=xy$.

The Taylor expansion of $f(x,y)$ is given by
\begin{equation}
f(x,y)=\sum_{n,m\geq 0}c_{nm}x^ny^m.
\end{equation}

It is conceptually convenient to decompose this sum into three parts:
\begin{equation}
\begin{split}
f_x(x)&=\sum_{n\geq 0}c_{n,0}x^n\\
f_y(y)&=\sum_{m\geq 1}c_{0,m}y^m\\
f_{xy}(x,y)&=xy\sum_{n,m\geq 1}c_{n,m}x^{n-1}y^{m-1}
\end{split}
\end{equation}

The contributions from $f_x$ and $f_y$ will have the periodicities of layers 1 and 2 respectively, and thus do not contribute moir\'e modes. However, the term $f_{xy}$ exhibits the same collection of Fourier modes as the simplest case of $f(x,y)=xy$. This immediately follows from the fact that if the Fourier modes of $g_i(x)$ are nonzero only on some lattice, then $[g_i(x)]^k$ has Fourier modes nonzero only on the same lattice.

Accordingly, \textit{any} function which combines the two layers nonlinearly (noting that nonlinear terms in $f_x$ and $f_y$ can be understood as linear terms with different $g_i$) will result in the same set of moir\'e frequencies, demonstrating that the derivation provided in Sec.~\ref{Sec:FreqMethod} is not specific to products only.

Moreover, even if the obvious physical quantity to measure combines linearly, subsequent nonlinearities in measurement can generate cross-terms. E.g., acoustic beats arise with moir\'e frequencies (up to a factor of two) because the human ear is sensitive to the square of pressure deviation rather than the absolute pressure. While pressure adds linearly, this square adds cross-terms to the quantities that are \textit{actually} measured (and is the origin of the factor of $2$ for pure cosines).

These nonlinearities can even occur beyond the measurement itself, instead arising at the visualization stage. E.g., a pattern that quickly oscillates between black and white may have the same average intensity as a solid gray (meaning the human eye can't tell the difference at a distance), but if instead plotted with color the moir\'e may manifest anyway. (Or, in fact, even in grayscale, due to nonlinear sensitivities of the human eye to light.)

\subsection{Moir\'e frequences away from $\Gamma$\label{Apx:QuantOp}}

Not every quantity of interest in physics shares the periodicity of the underlying lattice. For instance, in graphene, the low energy physics arises from the $K$ point, where the electron annihilation operators under translation by $R$ transform as $c_K^\dagger\rightarrow e^{-iK\cdot R}c_K^\dagger$ (and creation operators transform oppositely).

Accordingly, in this section, we consider products of terms that arise away from the origin of the BZ, such as the interlayer hopping term $t(c_{k,1}^\dagger c_{k,2}+h.c.)$ in twisted bilayer graphene. In the process, we also find which combinations of points in the original layers' BZs can couple to form moir\'e-periodic quantities.

Consider an observable $O=\Pi_i O_i(x)$ where $O_i$ lives at $k_i$ in the BZ of layer $i$, such that
\begin{equation}
    O_i(x+R)=e^{ik_i\cdot R}O_i(x)
\end{equation}
whenever $R$ is a lattice vector of layer $i$. Suppose also, for the moment, that $\sum_i k_i=0$. Then, since $\Pi_i e^{-ik_i\cdot x}=1$, $O(x)$ can also be written as
\begin{equation}
    O(x)=\Pi_i O_i(x)=\Pi_i O_i(x)e^{-ik_i\cdot x}.
\end{equation}

Each term $O_i(x)e^{-ik_i\cdot x}$ has the periodicity of layer $i$. Following the same argument as Sec.~\ref{Sec:FreqMethod}, $O(x)$ has a moir\'e periodicity when $\sum_i k_i = 0$.

If the condition $\sum_i k_i=0$ does not hold, then we use the sum $k_T=\sum_i k_i$ to give a phase prefactor
\begin{equation}
    O(x)=\Pi_i O_i(x)=e^{ik_T\cdot x}\Pi_i O_i(x)e^{-ik_i\cdot x}.
\end{equation}
This extra prefactor adds a periodicity of its own, which will generally (unless $k_T\approx 0$) ensure that the term $\Pi_i O_i(x)$ oscillates on a lengthscale much more rapid than the moir\'e lengthscale.

Therefore, this derivation directly illustrate which sets of points in the BZ result in moir\'e-scale physics when coupled, i.e., those whose momenta add to zero. Since these points do not lie on the reciprocal lattice, the arguments of the previous section about genericity of moir\'e periodicity beyond simple products \textit{do not apply}.

In TBG, this implies that the $K$-to-$K$ interlayer hopping term has moir\'e periodicity ($c_{K,1}c_{K,2}^\dagger$ has $\sum_i k_i=0$ at zero twist), whereas an analogous $K$-to-$K'$ interlayer hopping term does not ($c_{K,1}c_{K',2}^\dagger$ does not have $\sum_i k_i=0$).

\section{Near-commensurate moir\'e\label{Apx:NearComm}}
We here formalize the heuristic argument provided in Sec.~\ref{Sec:RSpCommens} as to the size of the moir\'e structure by explicitly defining the configuration space lattice (and the corresponding matrix of lattice vectors $A_\text{cs}$).

For each layer $i$, define the original lattice vectors as $A_{0,i}$ and the lattice vectors after a small deformation as $A_i$. The matrix describing the twist for layer $i$ is then given by $M_i=A_iA_{0,i}^{-1}$, analogous to Eq.~\eqref{eq:MiDef}. The derivation of the translation of each layer then proceeds exactly as follows that equation, reproducing Eq.~\eqref{eq:ConfigTranslation}.

We now refer to the formula for configuration space, Eq.~\eqref{eq:GeometricCSDef}, replicated here for convenience:
\begin{equation}
    T_\text{config}=T_1\times T_2/T_{12}.
\end{equation}
Recall $T_i$ is the space of translations of layer $i$, and $T_{12}$ is the simultaneous space of translations of both layers (all taken before twisting and modulo lattice translations of the respective lattices). The minor technical difficulty now is computing this particular group.

This is simplest if we write $T_i=\tau_i/L_i$, where $\tau_i$ denote the space of \textit{all} translations of layer $i$ (before twisting) - i.e., \textit{without} modding out by the lattice vectors - and $L_i$ are the translations of layer $i$ by its lattice vectors. (Similarly define $\tau_{12}$ as the simultaneous translation of both layers by arbitary amounts and $L_{12}$ as simultaneous translation by a commensurate lattice vector). The key to computing the group structure is then the identification
\begin{equation}
    \begin{split}
    T_\text{config}&=T_1\times T_2/T_{12}=\frac{(\tau_1/L_1)\times(\tau_2/L_2)}{\tau_{12}/L_{12}}\\
    &=\left(\frac{\tau_1\times\tau_2}{L_1\times L_2}\right)/\left(\frac{\tau_{12}}{L_{12}}\right)\\
    &=\frac{\tau_1\times\tau_2}{\tau_{12}\cdot(L_1\times L_2)}\\
    &=\left(\frac{\tau_1\times\tau_2}{\tau_{12}}\right)/\left(\frac{L_1\times L_2}{L_{12}}\right).
    \end{split}
\end{equation}
Note $\times$ denotes direct products and $\cdot$ group multiplication (i.e., performing the translations of layers sequentially). The only subtle aspect of this proof is the identification $L_{12}=\tau_{12}\cap(L_1\times L_2)$; the rest are properties of abelian groups.

We are now prepared to derive the analogue of Eq.~\eqref{eq:GeomConfigSpaceFormula}. The numerator of our product indicates that we are looking for relative translations of the two layers, as intuitively expected. The denominator tells us that we then have to mod out by \textit{both} layers' pre-twist lattices to find the configuration space, which is what makes the final result counterintuitive.

The commensurate version of Eq.~\eqref{eq:GeomConfigSpaceFormula} is therefore
\begin{equation}
    \tilde C(x)=(M_2^{-1}-M_1^{-1})x\mod \{A_{0,1}\ ,\ A_{0,2}\}
\end{equation}
where working modulo two lattices means that two elements are equivalent if they differ by any combination of lattice vectors of the two lattices. Note that if the two layers are initially identical, such that $A_{0,1}=A_{0,2}=A$, then this reduces to Eq.~\eqref{eq:GeomConfigSpaceFormula}, as expected. Otherwise, however, more care is needed to understand the consequences of working modulo two (commensurate) lattices.

To work modulo two lattices simultaneously is the same as working modulo their Minkowski sum. If the set of lattice vectors for layer $i$ is written as $L_i$, then the Minkowski sum of the two lattices is defined by
\begin{equation}
    L=\{v_1+v_2|v_1\in L_1,v_2\in L_2\}.
\end{equation}
This lattice can also be understood as the largest lattice containing both of the original lattices as subsets, and is perhaps most familiar as the new reciprocal lattice when one folds the Brillouin zone from a commensurate structure. (This construction is also called the DSC lattice; see, e.g., Ch. 13 of Ref.~\onlinecite{bollmann2012crystal}.)

The key to understanding the size of the moir\'e pattern, then, is that the lattice that defines the torus of configuration space (i.e., which replaces $A$ in Eq.~\eqref{eq:GeomConfigSpaceFormula}) is this Minkowski sum of the original lattices. Allowing $A_\text{cs}=A_{L_1+L_2}$, then, the formulas for configuration space and moir\'e lattice follow immediately as
\begin{equation}
    \tilde C(x)=(M_2^{-1}-M_1^{-1})x\mod A_\text{cs}
\end{equation}
and
\begin{equation}\label{eq:GeomConfigMoireLVsCS}
    A_M=(M_2^{-1}-M_1^{-1})^{-1}A_\text{cs}.
\end{equation}
A useful corollary of this last formula is the ability to derive the orientation of the moir\'e patterns as well as their size.

\subsection{Relative coordinate picture\label{Sec:CommensRelCoord}}
We now consider the alternative picture of configuration space in terms of relative coordinates, as defined in Eq.~\ref{eq:AlgebraicCSDef}, and generalize to this near-commensurate twisting case. The key complication is to correctly define the relative coordinates.

If one takes relative coordinates modulo the individual layers' lattice vectors, then the difference between the relative coordinates (per Eq.~\eqref{eq:AlgebraicCSDef}) will no longer be a slowly-varying function, hence it cannot be the quantity that is modulated by the long-wavelength moir\'e pattern. A natural first alternative is to take the relative coordinates in the commensurate cell instead.

Define $A_i$, $A_{0,i}$, and $A_C$ as the lattice vectors for the post-twist layer $i$ lattice, the pre-twist layer $i$ lattice, and the commensurate structure, respectively. The commensurate relative coordinates $x_i^C(x)$ can be defined as
\begin{equation}\label{eq:CommensRelCoords}
    x_i^C(x)=A_C^{-1}A_{0,i}A_i^{-1}x\quad\text{mod }\I.
\end{equation}
The set of commensurate relative coordinates corresponding to the lattice vectors of layer $i$ can then be written as $A_C^{-1}A_{0,i}$ mod $\I$. (Contrasting with \eqref{eq:relcoord}, we have three matrices instead of one. In this case, the $A_{0,i}A_i^{-1}$ part ``untwists" back to the physical coordinates in the pre-twist system, whereupon we can then understand the commensurate cell $A_C$. In the identical lattice case, the untwisting and the mapping onto the shared cell of the layers is the same step. In other words, previously, we effectively had $A^{-1}A_{0,i}A_i^{-1}$, but $A=A_{0,i}$, so the first two terms cancel.)

The key physical insight to understanding configuration space is that a relative translation between the layers by either layer's lattice vector preserves the configuration. E.g., a relative translation by a lattice vector of layer $1$ preserves the configuration; this is clearly necessary, since this relative translation can be interpreted as an absolute translation of layer $1$ by one of its lattice vectors, hence a symmetry of the system (and therefore necessarily the same configuration).

The lattice vectors of the individual layers thereby generate a collection of relative translations which are analogous to the additional symmetries in a magnetic space group: a relative translation by a fraction of a commensurate cell combined with an action on the internal degrees of freedom of each unit cell (here a permutation of sublattice labels of each layer rather than swapping electron spin). These symmetries are indexed by specifying a lattice vector for each layer that lies within the commensurate WS cell; a relative translation by the sum of those vectors is a non-symmorphic symmetry.

Working modulo these additional symmetries ultimately constitutes working modulo $A_C^{-1}A_{cs}$ for each layer (with $A_{cs}$ given by the Minkowski sum, as described in the previous section). Rather than working modulo these additional symmetries, we scale up by the inverse of this matrix, resulting in a configuration space in relative coordinates given by
\begin{align}
    \tilde{\tilde{C}}(x)&=A_m^{-1}(A_{0,2}A_2^{-1}-A_{0,1}A_1^{-1})x&&\text{mod }\I \label{eq:MinkCSFormula}\\
    &=x_2^\text{cs}(x)-x_1^\text{cs}(x)&&\text{mod }\I \label{eq:AlgebraicCSMinkDef}
\end{align}
where $x^\text{cs}=A_\text{cs}^{-1}A_{0,i}A_i^{-1}x\quad\text{mod }\I$. Therefore, the picture as the difference of relative coordinates still works, but the relative coordinates are in the lattice defined by the Minkowski sum of the original lattices.

To conclude, the moir\'e lattice vectors can therefore be computed again as
\begin{equation}\label{eq:MinkMLVFormula}
    A_M=(A_{0,2}A_2^{-1}-A_{0,1}A_1^{-1})^{-1}A_\text{cs},
\end{equation}
which is easily seen to be equivalent to Eq.~\eqref{eq:GeomConfigMoireLVsCS} (since $A_{0,i}A_i^{-1}=M_i^{-1}$).

\section{1D moir\'e in 2D systems\label{Apx:AsymmMoire}}

Thus far, we have been considering layers with high shared rotational symmetry (fourfold or sixfold). Without that symmetry, the moir\'e patterns can become more exotic.

The lower-symmetry cases have been studied in the case of two nearly-aligned layers, such as twisted rectangular lattices \cite{Kennes_2020,Wang_2022,Kariyado_2019,Fujimoto_2021}. In this case, the resulting moir\'e pattern also lacks the higher rotational symmetry. However, while the anisotropy may result in one direction visually dominating over the other (as can be seen in Fig.~1b of Ref.~\onlinecite{Wang_2022}), the resulting pattern is still periodic on a 2D lattice.%

However, if the layers are not identical, then they may have frequency vectors that align only along one axis. For instance, consider two rectangular lattices with the same lattice constant in the x-direction, but in the y-direction one of the lattices is larger by a factor of $\sqrt{2}$. Alternatively, consider stacking a hexagonal and square lattice with the same lattice constant.

In these cases, configuration space is difficult to work with because the resulting space is not a torus, but a cylinder, unbounded in one direction. However, the formalism in momentum space is more clear.

\begin{figure}
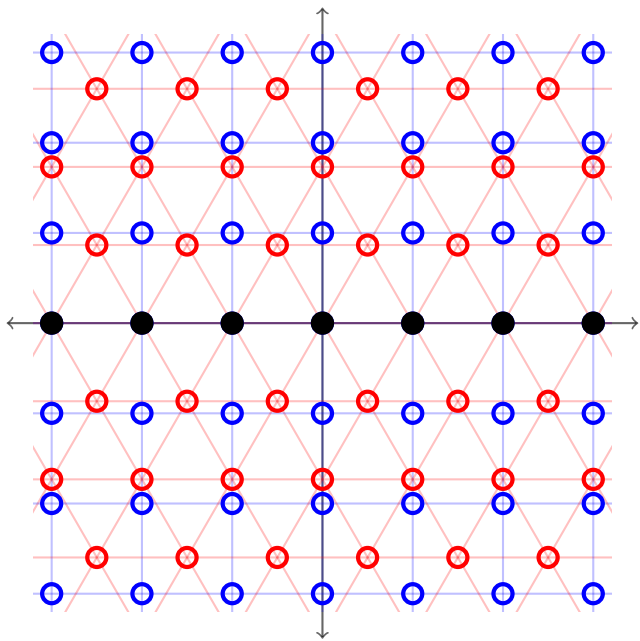

    \centering
    \include{Images_ReciprocalSpaceDiagrams_SqHexRecipSpace}
    \caption{Frequency modes of triangular (red circles) and square (blue circles) lattices with the same lattice constants at zero twist. The black filled circles indicate shared modes that yield moir\'e modes after a small twist or mismatch. Note since they only align along a 1D subspace, the resulting moir\'e is also 1D.}
    \label{fig:SqHexRecip}
\end{figure}

Consider the hexagonal-on-square setup for concreteness. As illustrated in Fig.~\ref{fig:SqHexRecip}, the two lattices will have one direction along which their frequency vectors agree. After a small twist, those frequency vectors will produce a low-frequency pattern in the orthogonal direction, resulting in a 1D moir\'e pattern.

\begin{figure}
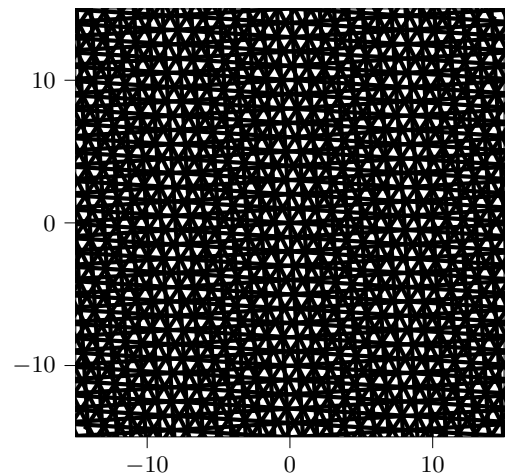

    \centering
    \include{Images_2DMoireTikZ_SqHex_SqHex}
    \caption{A unit square lattice on a unit triangular lattice at a relative twist angle of $7^\circ$ exhibits a 1D moir\'e pattern.}
    \label{fig:SqHex}
\end{figure}

However, there is no moir\'e pattern in the orthogonal direction. The result, to a first approximation, is a 1D set of stripes, rather than a 2D set of spots (see Fig.~\ref{fig:SqHex}). (A highly anisotropic 2D moir\'e pattern may appear instead if there is approximate commensuration in the other frequency direction.)

A 1D moir\'e has been considered in the case of uniaxial strain \cite{GrapheneStrainMoire1,Fu1DStrain,Bai_2020,Timmel_2020,Timmel_2021,Becker1DHofs}. However, setups like the triangle-on-square stacking illustrated in Fig.~\ref{fig:SqHex} are different than a uniaxial strain in that they are not fine-tuned. A 1D moir\'e only arises from uniaxial strain if it is \textit{purely} uniaxial; any small strain in the orthogonal direction will result in a 2D moir\'e, because the 1D moir\'e relies on having exactly-matched frequency vectors in the orthogonal direction. On the other hand, setups where the lattice frequency vectors only align in one direction form a 1D moir\'e for generic combinations of strain and twist near the singular structure.%

\section{Commensurability proof\label{Apx:CommensPf}}
In this appendix, we provide a short proof that for an N-layer system, if every layer is commensurate with layer 1, then there is an overall commensurate structure (assuming threefold or fourfold rotational symmetry). We begin with the argument for a trilayer system.

We will show that given a trilayer system where layers 1 and 2 are commensurate (with commensurate WS cell $C_{12}$ and lattice $L_{12}$) and layers 1 and 3 are commensurate (with commensurate WS cell $C_{13}$ and lattice $L_{13}$), there exists a commensurate unit cell for all three layers.

To prove this, consider the $C_{12}$ and the (finite) collection of points within that cell corresponding to lattice vectors of layer 1. All points in $L_{13}$ must map onto this (finite) set under the quotient map by $L_{12}$. Therefore, by pidgeonhole principle, there must be two points in $L_{13}$ that map to the same point in $C_{12}$. Their difference (in the original space), therefore, must be an element of $L_{12}$. Since it is also an element of $L_{13}$, it is a commensurate lattice vector of the whole system, showing that the system as a whole has commensurate lattice vectors. This completes the proof.

The general case follows by induction: take a commensurate unit cell for layers $1$-$N$, and enlarge to fit the $(1,N+1)$ unit cell in exactly the same way. The same argument easily generalizes to other combinations of layers being commensurate.

\section{Optimal singular structures\label{Apx:OptSingStruct}}
Suppose we have a stack of $N$ layers that generate a moir\'e pattern from lowest harmonics. Take the lattice vectors of layer $i$ to form the columns of the matrix $A_i$. For suitable choices of those lattice vectors, the moir\'e lattice vectors can be written
\begin{equation}
    A_M=(\sum_i A_i^{-1})^{-1}
\end{equation}

Take a nearby singular configuration to measure with respect to, with (corresponding) lattice vectors forming the columns of $B_i$; note the specific configuration is not determined by the $A_i$. The map $M_i=A_iB_i^{-1}$ then determines the relative translation of the lattice, as discussed in Sec.~\ref{Sec:SingStrucDegenConfigSpaceConnection}.

Suppose now that we choose our singular structure $B_i$ (near $A_i$) such that, for each layer $i$, $M_i$ satisfies the condition
\begin{equation}
    M_iA_M=A_iZ
\end{equation}
for $Z$ some integer matrix which satisfies the rotation symmetry of the original lattices.

Typically, if one ``undoes" the small deformation to identify the singular structure, the different equivalent points on the moir\'e pattern will differ by a nontrivial trivial transformation. In this case, however, the stackings will be identical, not identical up to those additional transformations. In other words, for any two moir\'e-equivalent points, deforming the lattices back to the special singular configuration centered about either point will yield the same quasicrystal structure (not just a physically equivalent one).

One can then potentially regard the line in between as a defect in the quasicrystalline singular structure, analogous to the description provided in Ref.~\onlinecite{KimTBGDefectNetwork}.

In the simple example case of TBG, the allowed transformations are an overall rescaling. The hypothetical goal of such a rescaling would be to cause the patterns of different AA regions, if untwisted and continued to a full lattice, to ``mesh" if one continued them. This essentially amounts to rendering the ``overall translation" degree of freedom $T_{12}$ described in Eq.~\ref{eq:GeometricCSDef} as being trivial on every equivalent moir\'e spot (relative to each other).

\iftrue


\section*{Supplementary material}

# Supplement 1: Moiré Patterns of 1D Systems

In this supplement, we examine moiré patterns in 1D lattices to illustrate how the moiré lattice is related to the commensurate lattice. We also illustrate intrinsically trilayer moiré in 1D for completeness. Notice that in 1D, there is not a notion of twisting, but moiré patterns can be obtained nonetheless by stacking lattices with different lattice constants.

The moiré patterns illustrated in this supplement are sensitive to display resolution; for optimal viewing, zooming the relevant figures to full screen width is recommended.

## S1-1  Near 1:1 matching

We begin with the simplest case of nearly-matched lattices, where one lattice is $1 + \delta$ times the other, with $\delta > 0$, and set the lattice constant of the smaller lattice to one. (The two-chain model this yields has been studied; see, e.g., [S1-1].)

We now derive the moiré scale in real space. From the characterization of moiré patterns given in Sec. II B 1, the moiré lattice is the set of points $x$ where the relative coordinates of the two unit cells are the same. The relative coordinate for the unit-length lattice is $x$ mod 1, and the relative coordinate for the $(1+\delta)$-length lattice is $\frac{x}{1+\delta}$ mod 1. The moiré lattice is therefore given by the set of $x$ where

$$x = \frac{x}{1+\delta} \mod 1. \tag{S1-1}$$

This simplifies to

$$(1+\delta)x = x \mod 1+\delta \tag{S1-2}$$
$$\delta x = 0 \mod 1+\delta, \tag{S1-3}$$

which is satisfied when $\delta x = k(1+\delta)$ for any integer $k$. Taking $k = 1$ gives the moiré lengthscale

$$x = \frac{1+\delta}{\delta} = 1 + \frac{1}{\delta}. \tag{S1-4}$$

In the case that $\delta = \frac{m}{n}$, it follows that the moiré scale is $\frac{n+m}{m}$, i.e., the moiré scale is $\frac{1}{m}$ of the commensurate scale. In the specific case where $m = 1$, the commensurate unit cell coincides with the moiré unit cell. This is illustrated in Fig. S1-1.

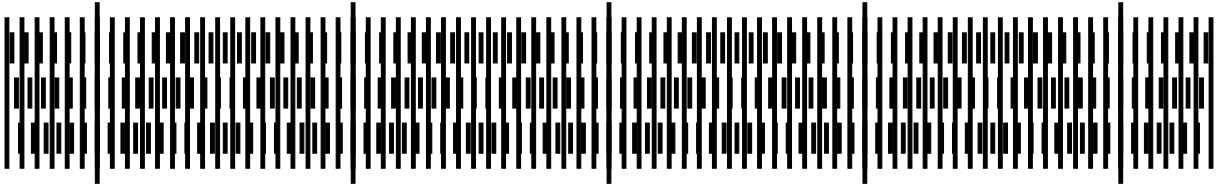

Figure S1-1: 1D moiré patterns illustrating commensurate and moiré unit cells. The relative lattice lengths are $\delta = -1/18, -2/19$, and $-3/20$, which all share a commensurate supercell (indicated by longer black lines). In the first row the commensurate and moiré unit cells coincide but in the second and third rows, the moiré pattern is one-half and one-third the commensurate unit cell, respectively.

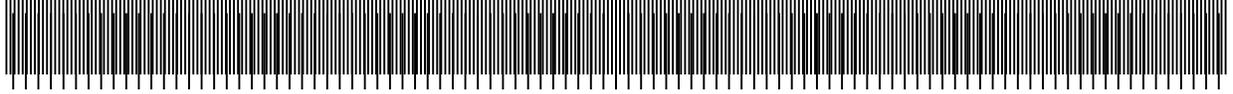

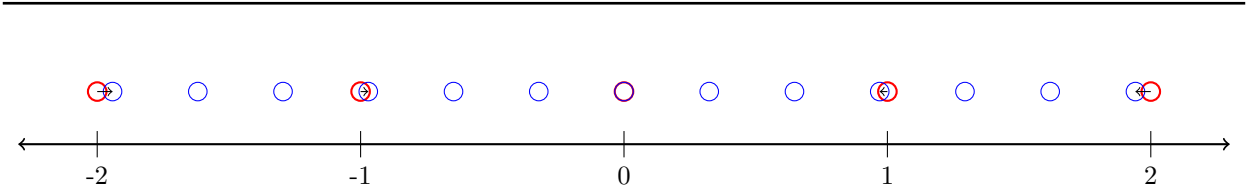

Figure S1-2: 1D moiré patterns near a 1:3 ratio, with $\delta = 0.03$. Top: moiré pattern from such a near-matching. Bottom: frequency modes of each layer (in red/blue) with arrows illustrating the resultant moiré frequencies.

Alternatively, the moiré unit cell can be deduced by the frequency picture described in Sec. III. The unit lattice has unit frequency, while the larger lattice has frequency $\frac{1}{1+\delta}$. Therefore, the moiré frequency is

$$1 - \frac{1}{1+\delta} = \frac{\delta}{1+\delta}. \tag{S1-5}$$

This is precisely the moiré lattice size given in Eq. (S1-4).

## S1-2 Near 1:p matching

We now consider the case where one lattice has a unit length unit cell and the other has a cell of size $p(1+\delta)$. To be concrete, we consider the case $p = 3$, with the moiré pattern illustrated in the top half of Fig. S1-2.

Using the frequency picture described in Sec. III, consider the $p$th mode of the $p$-length lattice and the first mode of the unit lattice, as illustrated in the bottom half of Fig. S1-2. These frequencies match at commensurability and therefore are slightly displaced after a mismatch by $\delta$. Their difference determines the moiré lattice size to be $1 + \frac{1}{\delta}$, the same size as in the 1:1 case.

We now characterize the same pattern in real space, following Appendix B 1. Instead of the moiré unit cell being determined by the set of points where the relative coordinates are equal, it is defined by the set of points where $p$ times the relative coordinate of the larger lattice equals the relative coordinate of the smaller lattice (mod 1). To derive this intuition, we reframe the 1:1 case in a way that will generalize.

In the 1:1 case, two points $x, y$ in the moiré pattern are equivalent if the difference between the relative coordinates of the two layers at $x$ is the same as at $y$, because then the two points have the same local stacking. Equivalently, consider the local stacking arrangement at point $x$ and then ask whether the relative coordinates from each layer at $y$ appear in the stacking at $x$. If so, $x$ and $y$ are equivalent points in the moiré pattern; otherwise, they are different.

Comparing the relative coordinates within the commensurate unit cell at two points naturally extends to the 1:$p$ case; however, this approach gives the wrong result for moiré pattern size. There are $p$ points in the commensurate unit cell that correspond to any particular point in the smaller unit cell (as illustrated for the 1:3 case in Fig. S1-3a). Given a particular commensurate stacking, points in the moiré pattern where the relative coordinates of the two layers appear in that stacking are $p$ times more common than points where the relative coordinates in the commensurate cell are the same. These extra points are, in fact, equivalent points on the moiré pattern; accounting for these extra equivalent points yields the correct condition on relative coordinates.

Using that condition, it is possible to compute moiré pattern size in the real space picture. At a particular point $x$, the relative coordinate of the unit lattice is $x$ mod 1, and the relative coordinate of the $p(1+\delta)$-length

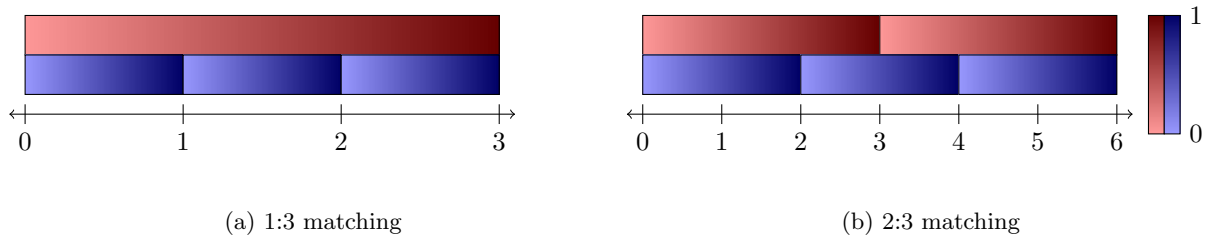

(a) 1:3 matching

(b) 2:3 matching

Figure S1-3: Commensurate unit cell for specified matchings. Shading indicates the relative coordinate in each corresponding layer. When one layer is stretched by a small amount $\delta$, the moiré lattice is determined by the set of points where the relative coordinates of each layer match one of the combinations shown here for the commensurate cell. For 1:3 matching, this condition is equivalent to Eq. (S1-6).

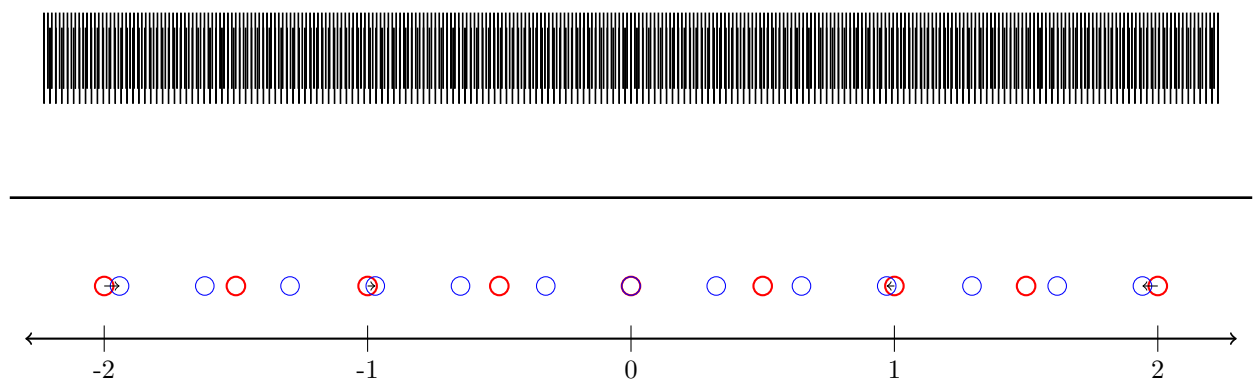

Figure S1-4: 1D moiré patterns near a 2:3 ratio, with $\delta = 0.01$. Above bar: moiré pattern from such a near-matching. Below bar: frequency modes of each layer (in red/blue), with arrows illustrating the resultant moiré frequencies.

lattice is $\frac{x}{p(1+\delta)}$ mod 1. Thus, the moiré lattice vectors occur at points $x$ satisfying

$$x = p\left(\frac{x}{p(1+\delta)}\right) \mod 1. \tag{S1-6}$$

We find the same moiré pattern size as in the 1:1 case, $1 + \frac{1}{\delta}$.

## S1-3 Near q:p matching

Now, we consider the commensurate structure with $q : p$ matching, where the first lattice has length $q$ and the second lattice has length $p(1+\delta)$. Without loss of generality, we consider $p$ and $q$ relatively prime, so that $\delta = 0$ yields a commensurate structure at $pq$.

In frequency space, the difference between the $p$th mode of the $p$-length lattice and the $q$th mode of the $q$-length lattice yields a moiré lattice size of $1 + \frac{1}{\delta}$ again. The case of $q = 2$ and $p = 3$ is shown in the lower half of Fig. S1-4. The moiré lattice is determined by $\gcd(p,q) = 1$. [I.e., the moiré lattice is $\sim \frac{1}{\delta}$, rather than $\frac{p}{\delta}$ or $\frac{q}{\delta}$, to leading order in $\delta$.] In the framework of Appendix B, this is because the Minkowski sum of two 1D lattices with lattice constants $a$, $b$ has lattice constant $\gcd(a,b)$.

In real space, the matching condition is equivalent to the 1:$p$ case: a moiré lattice vector occurs at any position where the relative coordinate of the top and bottom layer match any combination that would occur in the commensurate cell. The 2:3 case is illustrated in Fig. S1-3b. This definition yields the same moiré length of $1 + \frac{1}{\delta}$.

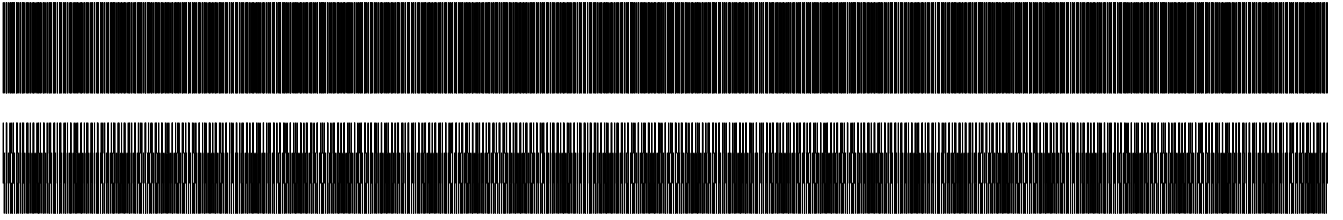

Figure S1-5: Intrinsically trilayer 1D moiré patterns. A 1:$\phi$:$(\phi-1)$ ratio of frequencies, where $\phi = \frac{1+\sqrt{5}}{2}$ is the golden ratio, produces a singular structure. We deviate from that singular structure by varying the first lattice constant by $\delta = 0.02$. The top image overlaps all three layers, showing a periodic structure with approximately eight periods in the displayed range. (A smaller bilayer moiré pattern is also visible.) The bottom shows the individual pairs of bilayers (from top to bottom: 1:$(\phi-1)$, 1:$\phi$, $\phi$:$(\phi-1)$), which do not exhibit any moiré patterns on the longer lengthscale.

## S1-4   Trilayer moiré

Intrinsically trilayer moiré is also possible in 1D. In Fig. S1-5, we illustrate a moiré pattern between three layers with periods $1 + \delta$, $\phi$, and $\phi - 1$, where $\phi = \frac{1+\sqrt{5}}{2}$ is the golden ratio. The singular structure at $\delta = 0$ results because $\frac{1}{\phi} = \phi - 1$. In addition to the intrinsically trilayer pattern, there are bilayer moiré patterns that are also visible in the figure.

## References

[S1-1] Qiang Gao and Eslam Khalaf. Symmetry origin of lattice vibration modes in twisted multilayer graphene: Phasons versus moiré phonons. *Phys. Rev. B*, 106:075420, Aug 2022.

# Supplement 2: Raw Moiré Images

In this supplement, we re-present several figures from the main text without the guides to the eye so that the visual arrangements can be viewed more objectively. These are presented as Figs. S2-1, S2-2, and S2-3.

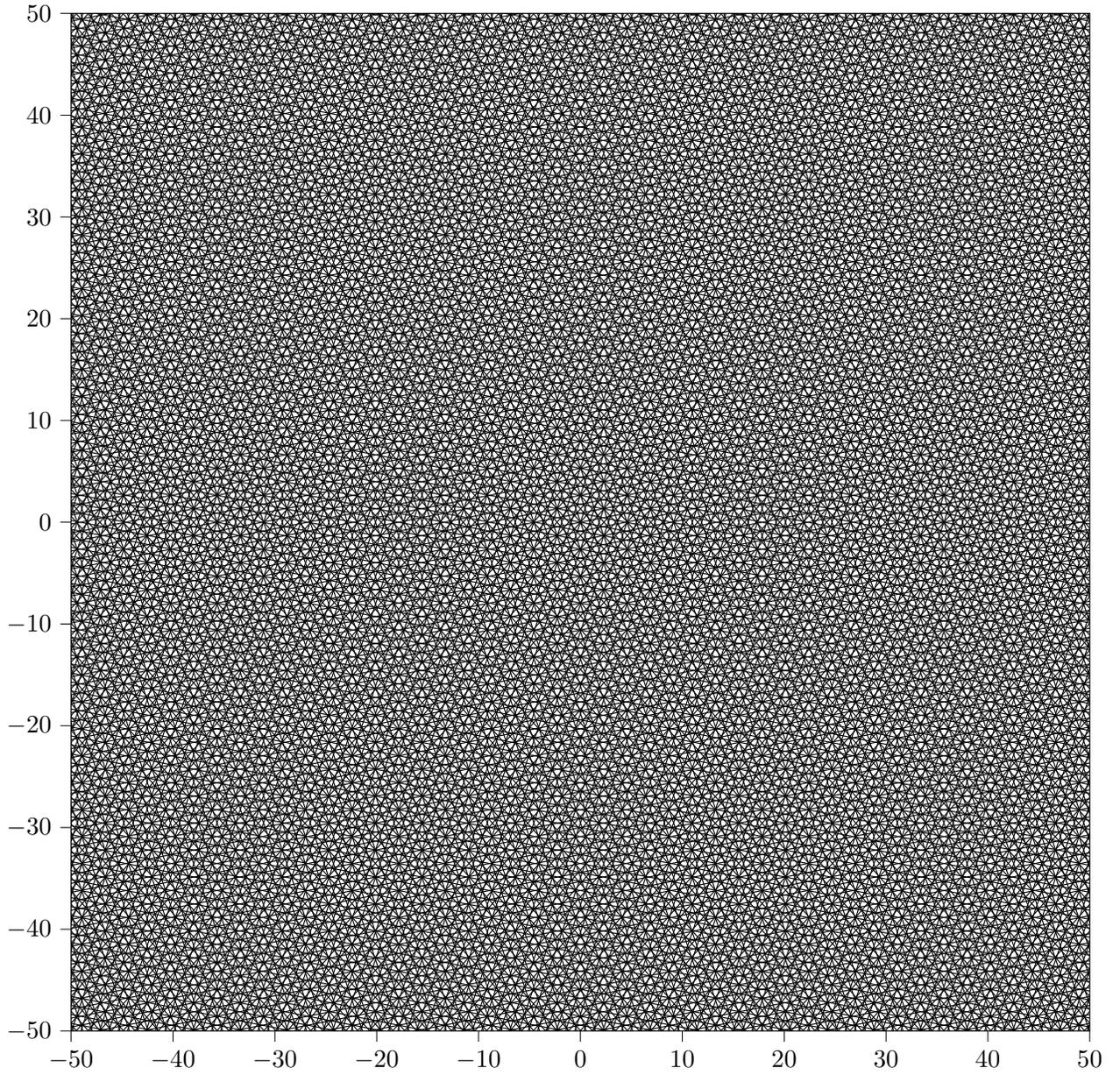

Figure S2-1: A larger and unannotated illustration of Fig. 4, showing two unit square lattices at a relative twist of 0.6° away from the 36.9° commensurate angle.

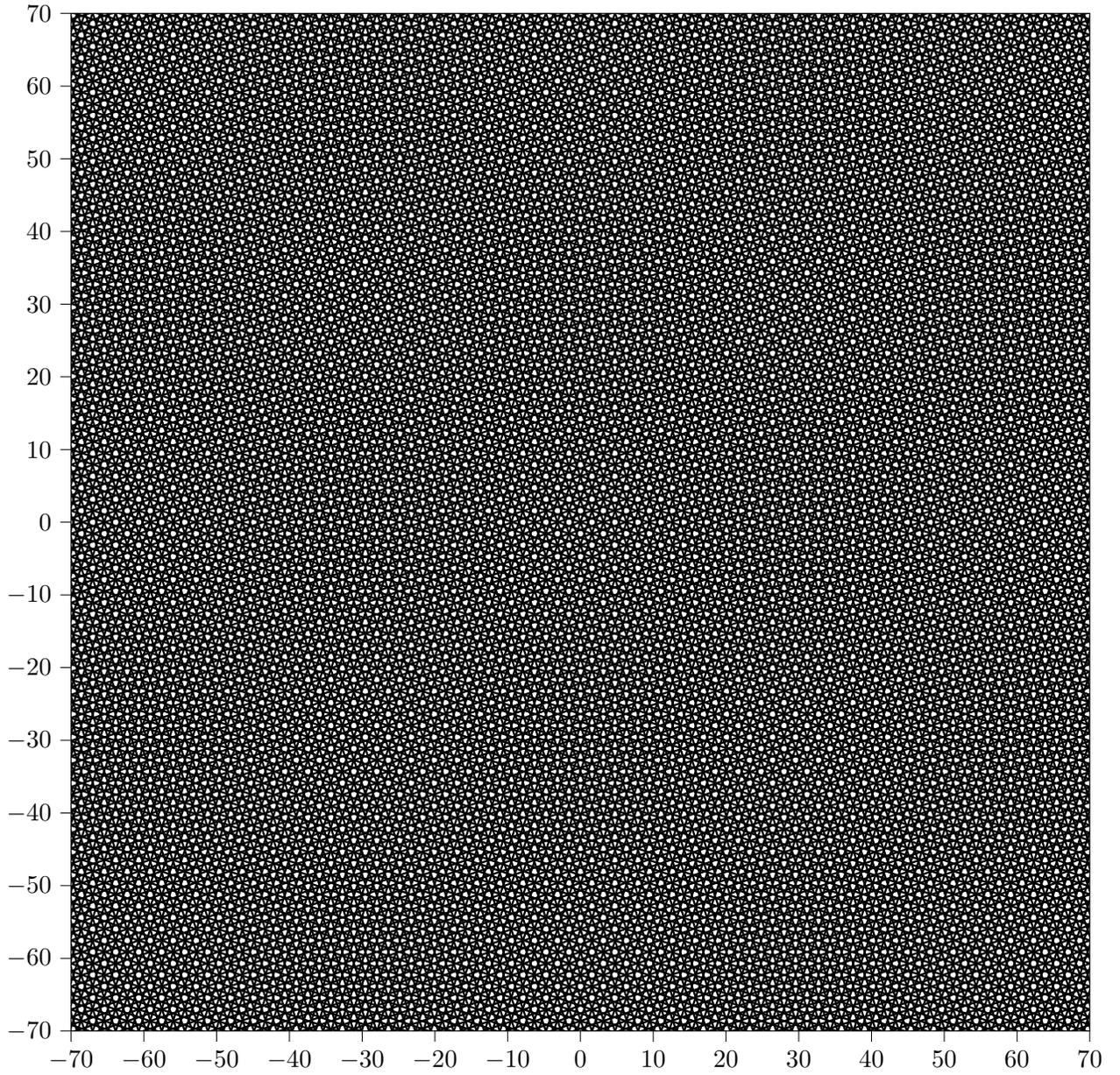

Figure S2-2: A larger and unannotated illustration of Fig. 5, showing two unit square lattices at a relative twist of 0.6° away from the 36.9° commensurate angle.

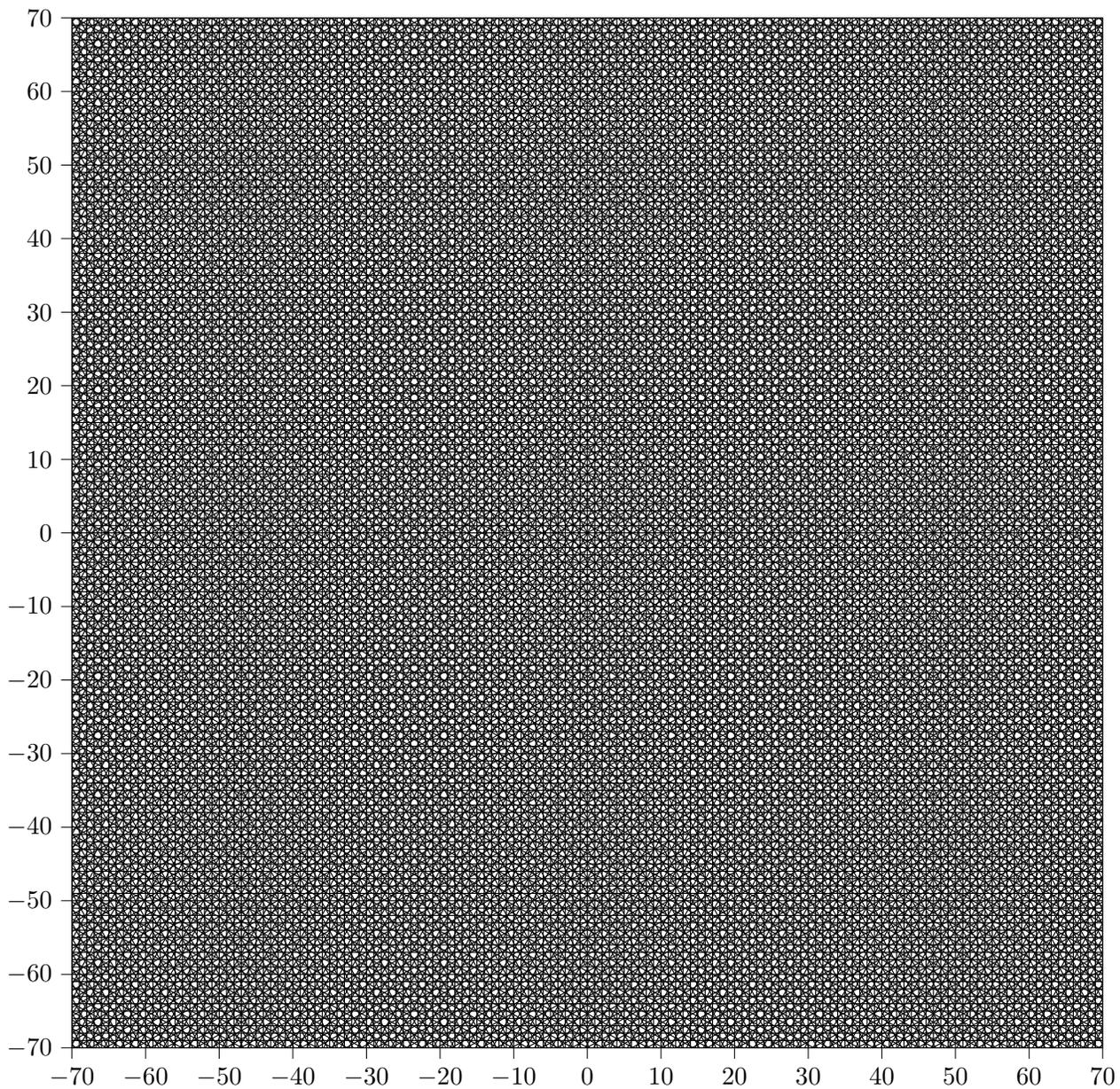

Figure S2-3: A larger and unannotated illustration of Fig. 8, showing three unit square lattices near 60.7°.


\fi
\end{document}

%% file: Images_TableImages_TBLGSchematic.tex
\begin{tikzpicture}
\begin{scope}[red,rotate=7]
    \draw (0:1) -- (60:1) -- (120:1) -- (180:1) -- (240:1) -- (300:1) -- cycle;
\end{scope}
\begin{scope}[blue,rotate=-7]
    \draw (0:1) -- (60:1) -- (120:1) -- (180:1) -- (240:1) -- (300:1) -- cycle;
\end{scope}
\end{tikzpicture}

%% file: Images_TableImages_CommensTBLGSchematic.tex
\begin{tikzpicture}[scale=0.2]
\tikzmath{
    \commsize=sqrt(7)*sqrt(3);
}
\def\hex{\draw (0:1) -- (60:1) -- (120:1) -- (180:1) -- (240:1) -- (300:1) -- cycle;}
\def\shiftedhex{\begin{scope}[shift={(0:2)}]\begin{scope}[shift={(60:2)}]\hex\end{scope}\end{scope}}
\def\hexcommenscell{
    \hex
    \shiftedhex
    \begin{scope}[rotate=60]\shiftedhex\end{scope}
    \begin{scope}[rotate=120]\shiftedhex\end{scope}
    \begin{scope}[rotate=180]\shiftedhex\end{scope}
    \begin{scope}[rotate=240]\shiftedhex\end{scope}
    \begin{scope}[rotate=300]\shiftedhex\end{scope}
}
\def\shiftedhexcommenscell{
    \begin{scope}[shift={(0:1)}]\begin{scope}[shift={(60:1)}]\hexcommenscell\end{scope}\end{scope}
}
\def\hexdoublecommenscell{
    \hexcommenscell
    \shiftedhexcommenscell
    \begin{scope}[rotate=60]\shiftedhexcommenscell\end{scope}
    \begin{scope}[rotate=120]\shiftedhexcommenscell\end{scope}
    \begin{scope}[rotate=180]\shiftedhexcommenscell\end{scope}
    \begin{scope}[rotate=240]\shiftedhexcommenscell\end{scope}
    \begin{scope}[rotate=300]\shiftedhexcommenscell\end{scope}
}

\begin{scope}[thick,dotted, black!80!white]
    \draw (0,0) -- (0:\commsize);
    \draw (0,0) -- (60:\commsize);
    \draw (0,0) -- (120:\commsize);
    \draw (0,0) -- (180:\commsize);
    \draw (0,0) -- (240:\commsize);
    \draw (0,0) -- (300:\commsize);
    \draw (0:\commsize) -- (60:\commsize) -- (120:\commsize) -- (180:\commsize) -- (240:\commsize) -- (300:\commsize) -- cycle;
\end{scope}

\begin{scope}[rotate=12,red]
    \hexdoublecommenscell
\end{scope}
\begin{scope}[rotate=-12,blue]
    \hexdoublecommenscell
\end{scope}

\end{tikzpicture}

%% file: Images_TableImages_TTLGSchematic.tex
\begin{tikzpicture}
\begin{scope}[red,rotate=7]
    \draw (0:1) -- (60:1) -- (120:1) -- (180:1) -- (240:1) -- (300:1) -- cycle;
\end{scope}
\begin{scope}[blue,rotate=-7]
    \draw (0:1) -- (60:1) -- (120:1) -- (180:1) -- (240:1) -- (300:1) -- cycle;
\end{scope}
\begin{scope}[green!40!black]
    \draw (0:1) -- (60:1) -- (120:1) -- (180:1) -- (240:1) -- (300:1) -- cycle;
\end{scope}
\end{tikzpicture}

%% file: Images_TableImages_IntrinsicallyTrilayerSchematic.tex
\begin{tikzpicture}[scale=0.7]
\begin{scope}[red,rotate=60]
    \draw (-1,-1) grid (1,1);
\end{scope}
\begin{scope}[blue,rotate=-60]
    \draw (-1,-1) grid (1,1);
\end{scope}
\begin{scope}[green!40!black]
    \draw (-1,-1) grid (1,1);
\end{scope}
\end{tikzpicture}

%% file: Images_2DMoireTikZ_SqHD_SqSimple.tex
\begin{tikzpicture}

\definecolor{darkgray176}{RGB}{176,176,176}

\begin{axis}[
axis equal image,
point meta max=1,
point meta min=0,
tick align=outside,
tick pos=left,
x grid style={darkgray176},
xmin=-15, xmax=15,
xtick style={color=black},
y grid style={darkgray176},
ymin=-15, ymax=15,
ytick style={color=black}
]

\begin{scope}[]
    \begin{scope}[ultra thick]
        \begin{scope}[rotate around={3.366460663429805:(axis cs:0,0)}]
            \foreach \x in {-20,...,20} {
                \edef\temp{
                    \noexpand\draw(axis cs:-20,\x) -- (axis cs:20,\x);
                }
                \temp
                \begin{scope}[rotate around={90:(axis cs:0,0)}]
                    \temp
                \end{scope}
            }
        \end{scope}
        \begin{scope}[rotate around={-3.366460663429805:(axis cs:0,0)}]
            \foreach \x in {-20,...,20} {
                \edef\temp{
                    \noexpand\draw(axis cs:-20,\x) -- (axis cs:20,\x);
                }
                \temp
                \begin{scope}[rotate around={90:(axis cs:0,0)}]
                    \temp
                \end{scope}
            }
        \end{scope}
    \end{scope}
    \begin{scope}[rotate around={45:(axis cs:0,0)},color=red]
        \foreach \x in {-12.041594578792285,0,12.041594578792285} {
            \edef\temp{
                \noexpand\draw(axis cs:-20,\x) -- (axis cs:20,\x);
            }
            \temp
            \begin{scope}[rotate around={90:(axis cs:0,0)}]
                \temp
            \end{scope}
        }
    \end{scope}
    \begin{scope}[color=blue,thick]
        \foreach \x in {-17.029386365926385,-8.5146931829631925,...,17.029386365926385} {
            \edef\temp{
                \noexpand\draw(axis cs:-20,\x) -- (axis cs:20,\x);
            }
            \temp
            \begin{scope}[rotate around={90:(axis cs:0,0)}]
                \temp
            \end{scope}
        }
    \end{scope}
\end{scope}

\end{axis}
\end{tikzpicture}

%% file: Images_2DMoireTikZ_0DegTBLG_5DegTBLG.tex
\begin{tikzpicture}

\definecolor{darkgray176}{RGB}{176,176,176}

\begin{axis}[
axis equal image,
point meta max=1,
point meta min=0,
tick align=outside,
tick pos=left,
x grid style={darkgray176},
xmin=-30, xmax=30,
xtick style={color=black},xtick={-30,-15,0,15,30},
y grid style={darkgray176},
ymin=-30, ymax=30,
ytick style={color=black},ytick={-30,-15,0,15,30}
]

\begin{scope}[thick, dash pattern=on 2.7pt off 5.4pt]
    \begin{scope}[rotate around={2.5:(axis cs:0,0)}]
        \foreach \x in {-44.167295593006365,-42.43524478543749,...,44.167295593006365} {
            \edef\temp{
                \noexpand\draw(axis cs:-0.5,\x) -- (axis cs:40.5,\x);
                \noexpand\draw(axis cs:0.5,\x) -- (axis cs:-40.5,\x);
                \noexpand\draw[yshift=2.338268590218pt](axis cs:1,\x) -- (axis cs:41,\x);
                \noexpand\draw[yshift=2.338268590218pt](axis cs:-1,\x) -- (axis cs:-41,\x);
            }
            \temp
            \begin{scope}[rotate around={60:(axis cs:0,0)}]
                \temp
            \end{scope}
            \begin{scope}[rotate around={120:(axis cs:0,0)}]
                \temp
            \end{scope}
        }
    \end{scope}
    \begin{scope}[rotate around={-2.5:(axis cs:0,0)}]
        \foreach \x in {-44.167295593006365,-42.43524478543749,...,44.167295593006365} {
            \edef\temp{
                \noexpand\draw(axis cs:-0.5,\x) -- (axis cs:40.5,\x);
                \noexpand\draw(axis cs:0.5,\x) -- (axis cs:-40.5,\x);
                \noexpand\draw[yshift=2.338268590218pt](axis cs:1,\x) -- (axis cs:41,\x);
                \noexpand\draw[yshift=2.338268590218pt](axis cs:-1,\x) -- (axis cs:-41,\x);
            }
            \temp
            \begin{scope}[rotate around={60:(axis cs:0,0)}]
                \temp
            \end{scope}
            \begin{scope}[rotate around={120:(axis cs:0,0)}]
                \temp
            \end{scope}
        }
    \end{scope}
\end{scope}

\draw[red,very thick] (axis cs:0,0) circle (12pt);
\draw[blue,very thick] (axis cs:0,11.463) circle (10pt);

\end{axis}
\end{tikzpicture}

%% file: Images_2DMoireTikZ_SqbySqrt2_SqbySqrt2.tex
\begin{tikzpicture}

\definecolor{darkgray176}{RGB}{176,176,176}

\begin{axis}[
axis equal image,
point meta max=1,
point meta min=0,
tick align=outside,
tick pos=left,
x grid style={darkgray176},
xmin=-30, xmax=30,
xtick style={color=black},
y grid style={darkgray176},
ymin=-30, ymax=30,
ytick style={color=black}
]

\begin{scope}[very thick]
    \begin{scope}[rotate around={1.5:(axis cs:0,0)}]
        \foreach \x in {-40,...,40} {
            \edef\temp{
                \noexpand\draw(axis cs:-40,\x) -- (axis cs:40,\x);
            }
            \temp
            \begin{scope}[rotate around={90:(axis cs:0,0)}]
                \temp
            \end{scope}
        }
    \end{scope}
    \begin{scope}[rotate around={43.5:(axis cs:0,0)},scale=1.4142135623730951]
        \foreach \x in {-40,...,40} {
            \edef\temp{
                \noexpand\draw(axis cs:-40,\x) -- (axis cs:40,\x);
            }
            \temp
            \begin{scope}[rotate around={90:(axis cs:0,0)}]
                \temp
            \end{scope}
        }
    \end{scope}
\end{scope}

\end{axis}
\end{tikzpicture}

%% file: Images_ReciprocalSpaceDiagrams_36.9RecipSpace.tex
\begin{tikzpicture}[scale=1.0]
\begin{scope}[<->,draw opacity=0.6,thick]
    \draw (-4,0) -- (4,0);
    \draw (0,-4) -- (0,4);
\end{scope}
\draw[very thin,rotate={45},gray,draw opacity=0.25,scale={1/sqrt(5)}] (-7,-7) grid (7,7);
\draw[very thin,rotate={atan(3/4)/2},red,draw opacity=0.5] (-3,-3) grid (3,3);
\draw[very thin,rotate={-atan(3/4)/2},blue,draw opacity=0.5] (-3,-3) grid (3,3);

\foreach \x in {-3,...,3} {
    \foreach \y in {-3,...,3} {
        \edef\temp{
            \noexpand\draw (\x,\y) circle (3 pt);
        }
        \begin{scope}[rotate={atan(3/4)/2},red]
            \temp
        \end{scope}
        \begin{scope}[rotate={-atan(3/4)/2},blue]
            \temp
        \end{scope}
    }
}
\foreach \x in {-1,...,1} {
    \foreach \y in {-1,...,1} {
        \edef\temptwo{
            \noexpand\filldraw ({sqrt(5)*\x},{sqrt(5)*\y}) circle (3 pt);
        }
        \begin{scope}[rotate=45]
            \temptwo
        \end{scope}
    }
}

\begin{scope}[rotate={atan(3/4)/2},red,ultra thick]
    \draw (0,0) -- (2,0) -- (2,1);
\end{scope}
\begin{scope}[rotate={-atan(3/4)/2},blue,ultra thick]
    \draw (0,0) -- (1,0) -- (1,2);
\end{scope}

\end{tikzpicture}

%% file: Images_ReciprocalSpaceDiagrams_0RecipSpace.tex
\begin{tikzpicture}[scale=1.5,auto]
\begin{scope}[<->,draw opacity=0.6,thick]
    \draw (-2.5,0) -- (2.5,0);
    \draw (0,-2.5) -- (0,2.5);
\end{scope}
\draw[very thin,rotate=7,red,draw opacity=0.5] (-2,-2) grid (2,2);
\draw[very thin,rotate=-7,blue,draw opacity=0.5] (-2,-2) grid (2,2);
\foreach \x in {-2,...,2} {
    \foreach \y in {-2,...,2} {
        \edef\temp{
            \noexpand\filldraw (\x,\y) circle (1.5 pt);
        }
        \begin{scope}[rotate=7,red]
            \temp
        \end{scope}
        \begin{scope}[rotate=-7,blue]
            \temp
        \end{scope}
    }
}

\draw[very thick,->,red] (0,0) -- node{$k_1$} (7:1);
\draw[very thick,->,blue] (0,0) -- node[swap]{$k_2$} (-7:1);
\draw[very thick,->] (-7:1) -- node[swap]{$k_M$} (7:1);

\end{tikzpicture}

%% file: Images_Other_TriangleSingularStructures.tex
\begin{tikzpicture}[->,ultra thick,auto,swap,scale=1.5]
\coordinate (avoidcutoff) at (0,-0.1);

\draw[red] (0,0) -- node{$G_1$} (0:1);
\draw[green!40!black] (0:1) -- node{$G_2$} (60:1);
\draw[blue] (60:1) -- node{$G_3$} (0,0);

\draw[red] (1.5,0) -- node{$G_1$} +(0:1);
\draw[green!40!black] (2.5,0) -- node{$G_2$} +(140:1);
\draw[blue] (1.5,0)++(40:1) -- node{$G_3$} (1.5,0);

\draw[red] (3,0) -- node{$G_1$} +(0:1);
\draw[green!40!black] (4,0) -- node{$G_2$} +(140:0.653);
\draw[blue] (3,0)++(40:0.653) -- node{$G_3$} (3,0);

\end{tikzpicture}

%% file: Images_Other_TTLGSingStruct.tex
\begin{tikzpicture}[scale=1.5,<->,ultra thick]
\draw[opacity=0] (-2.2,-1.1) rectangle (1.3,1.5);

\draw[red] (-1,0)--(1,0);
\draw[blue] (0,-1)--(0,1);
\draw[red!50!blue] (-0.7,0.7)--(0.7,-0.7);
\draw[dashed, green!40!black] (-0.7,-0.7)--(0.7,0.7);

\node at (1,0)[label={above:{\large $\theta_{12}$}}] {};
\node at (0,1)[label={above:{\large $\theta_{23}$}}] {};
\node at (0,1)[label={[blue]left:{ (1,0,-1,0,0,0)}}] {};
\node at (-1,0)[label={[red]below:{(0,0,1,0,-1,0)}}] {};
\node at (-0.7,0.7)[label={[red!50!blue]left:{(1,0,0,0,-1,0)}}] {};
\end{tikzpicture}

%% file: Images_Other_TTLG121SingStruct.tex
\begin{tikzpicture}[scale=1.5,<->,ultra thick]
\draw[opacity=0] (-1.2,-1.1) rectangle (2.3,1.5);

\begin{scope}[gray,opacity=0.5]
    \draw (-1,0)--(1,0);
    \draw (0,-1)--(0,1);
\end{scope}
\node at (1,0)[label={above:{\large $\delta_{12}=\delta_{32}$}}] {};
\node at (0,1)[label={above:{\large $\theta_{12}=\theta_{23}$}}] {};
\begin{scope}[green!40!black]
    \draw[dashed] (0,-1)--(0,1);
    \draw[rotate=-90] (-1,0.3) parabola bend (0,0) (1,0.3);
    \node at (0.3,1)[label={right:{$(1,0,-2,0,1,0)$}}] {};
\end{scope}
\end{tikzpicture}

%% file: Images_Other_TwoKiteSingularStructures.tex
\begin{tikzpicture}[scale=1.,->,ultra thick,auto,swap]

\def\xshiftparam{4}
\def\yshiftparam{-2.5}

\coordinate (node0) at (0,0);
\coordinate (node1) at (1,-1);
\coordinate (node2) at (3,0);
\coordinate (node3) at (1,1);

\coordinate (node4) at ($(0,0)+(\xshiftparam,0)$);
\coordinate (node5) at ($(-60:1.414)+(\xshiftparam,0)$);
\coordinate (node6) at ($(1.871+0.707,0)+(\xshiftparam,0)$);
\coordinate (node7) at ($(60:1.414)+(\xshiftparam,0)$);

\draw (node0)--node{$G_1$}(node1);
\draw (node1)--node{$G_2$}(node2);
\draw (node2)--node{$G_3$}(node3);
\draw (node3)--node{$G_4$}(node0);

\draw (node4)--node{$G_1$}(node5);
\draw (node5)--node{$G_2$}(node6);
\draw (node6)--node{$G_3$}(node7);
\draw (node7)--node{$G_4$}(node4);

\end{tikzpicture}

%% file: Images_ExpSetups_TwoPotentialGraphene.tex
\begin{tikzpicture}[ultra thick,auto,scale=1.5]

\draw[thick] (0:1) -- (60:1) -- (120:1) -- (180:1) -- (240:1) -- (300:1) -- cycle;

\coordinate (G1end) at (-0.3,-0.2);
\draw[->,blue] (0:1) -- node[swap]{$G_1$} (G1end);
\begin{scope}[rotate around={-10:(G1end)}]
    \coordinate (G2end) at (0:1);
\end{scope}
\draw[->,blue] (G1end) -- node[swap]{$G_2$} (G2end);
\draw[->,red] (0:1) -- node{$G_M$} (G2end);

\begin{scope}[xshift=3cm]
    \draw[thick] (0:1) -- (60:1) -- (120:1) -- (180:1) -- (240:1) -- (300:1) -- cycle;
    \begin{scope}[rotate around={-10:(0:1)}]
        \coordinate (G1end) at (120:0.3);
        \draw[->,blue] (0:1) -- node[swap]{$G_1$} (G1end);
        \begin{scope}[rotate around={5:(G1end)}]
            \coordinate (G2end) at (240:1);
            \draw[->,blue] (G1end) -- node[swap]{$G_2$} (G2end);
        \end{scope}
    \end{scope}
    \draw[->,red] (240:1) -- node{$G_M$} (G2end);
    \draw[->,dashed] (0:1) -- node{$G_3$} (240:1);
\end{scope}
\end{tikzpicture}

%% file: Images_ReciprocalSpaceDiagrams_SqHexRecipSpace.tex
\begin{tikzpicture}[scale=1.2, ultra thick]

\begin{scope}[<->,draw opacity=0.6,thick]
    \draw (-3.5,0) -- (3.5,0);
    \draw (0,-3.5) -- (0,3.5);
\end{scope}

\begin{scope}[red,opacity=0.25,thick]
    \clip (-3.2,-3.2) rectangle (3.2,3.2);
    \foreach \x in {-4,...,4} {
        \begin{scope}[red]
        \draw (\x,0) -- ++ (60:5);
        \draw (\x,0) -- ++ (60:-5);
        \begin{scope}[rotate=60]
            \draw (\x,0) -- ++ (60:5);
            \draw (\x,0) -- ++ (60:-5);
        \end{scope}
        \begin{scope}[rotate=-60]
            \draw (\x,0) -- ++ (60:5);
            \draw (\x,0) -- ++ (60:-5);
        \end{scope}
        \end{scope}
        
        \begin{scope}[blue]
        \draw (\x,-4) -- (\x,4);
        \draw (-4,\x) -- (4,\x);
        \end{scope}
    }
\end{scope}

\begin{scope}[red]
\foreach \x in {-2.5,-1.5,...,2.5} {
    \draw (\x,0.8660254037844386) circle (3 pt);
    \draw (\x,2.598076211353316) circle (3 pt);
    \draw (\x,-0.8660254037844386) circle (3 pt);
    \draw (\x,-2.598076211353316) circle (3 pt);
}
\end{scope}

\foreach \x in {-3,...,3} {
    \draw[red] (\x,1.7320508075688772) circle (3 pt);
    \draw[red] (\x,-1.7320508075688772) circle (3 pt);
    \foreach \y in {-3,...,3} {
        \draw[blue] (\x,\y) circle (3 pt);
    }
    \filldraw[black] (\x,0) circle (3 pt);
}
\end{tikzpicture}

%% file: Images_2DMoireTikZ_SqHex_SqHex.tex
\begin{tikzpicture}

\definecolor{darkgray176}{RGB}{176,176,176}

\begin{axis}[
axis equal image,
point meta max=1,
point meta min=0,
tick align=outside,
tick pos=left,
x grid style={darkgray176},
xmin=-15, xmax=15,
xtick style={color=black},
y grid style={darkgray176},
ymin=-15, ymax=15,
ytick style={color=black}
]

\begin{scope}[]
    \begin{scope}[ultra thick]
        \begin{scope}[rotate around={3.5:(axis cs:0,0)}]
            \foreach \x in {-20,...,20} {
                \edef\temp{
                    \noexpand\draw(axis cs:-20,\x) -- (axis cs:20,\x);
                }
                \temp
                \begin{scope}[rotate around={90:(axis cs:0,0)}]
                    \temp
                \end{scope}
            }
        \end{scope}
        \begin{scope}[rotate around={-3.5:(axis cs:0,0)}]
            \foreach \x in {-20,...,20} {
                \edef\temp{
                    \noexpand\draw(axis cs:-20,\x) -- (axis cs:20,\x);
                }
                \temp
                \begin{scope}[rotate around={60:(axis cs:0,0)}]
                    \temp
                \end{scope}
                \begin{scope}[rotate around={120:(axis cs:0,0)}]
                    \temp
                \end{scope}
            }
        \end{scope}
    \end{scope}
\end{scope}

\end{axis}
\end{tikzpicture}